\documentclass[12pt]{article}
\newcommand{\secn}[1]{Section~\ref{#1}}
\newcommand{\bra}[1]{\langle{#1}|}
\newcommand{\ket}[1]{|{#1}\rangle}
\newcommand{\braket}[2]{\langle{#1}|{#2}\rangle}

\newcommand{\eq}[1]{Eq.~(\ref{#1})}

\def\beq{\begin{equation}}
\def\eeq{\end{equation}}
\def\beqa{\begin{eqnarray}}
\def\eeqa{\end{eqnarray}}
\newcommand{\sect}[1]{\setcounter{equation}{0}\section{#1}}
\renewcommand{\theequation}{\thesection.\arabic{equation}}
\newcommand{\EQ}{\begin{equation}}
\newcommand{\EN}{\end{equation}}
\newcommand{\bea}{\begin{eqnarray}}
\newcommand{\ena}{\end{eqnarray}}

\renewcommand{\a}{\alpha}
\renewcommand{\b}{\beta}

\newcommand{\shalf}{\frac{1}{2}}

\newcommand{\NP}[1]{Nucl.\ Phys.\ {\bf #1}}
\newcommand{\PL}[1]{Phys.\ Lett.\ {\bf #1}}

\newcommand{\PR}[1]{Phys.\ Rev.\ {\bf #1}}

\renewcommand{\thefootnote}{\fnsymbol{footnote}}
\def\one{{\hbox{ 1\kern-.8mm l}}}

\def\R{{\rm R}}
\def\ii{{\rm i}}

\newlength{\bredde}
\def\slash#1{\settowidth{\bredde}{$#1$}\ifmmode\,\raisebox{.15ex}{/}
\hspace*{-\bredde} #1\else$\,\raisebox{.15ex}{/}\hspace*{-\bredde} #1$\fi}

\begin{document}
\begin{titlepage}
\rightline{DFTT 33/99}
\rightline{NORDITA-1999/34 HE}
\rightline{\hfill June 1999}
\vskip 1.2cm
\centerline{\Large \bf (F,D$p$) bound states from the boundary state
\footnote{Work 
partially supported by the European Commission
TMR programme ERBFMRX-CT96-0045 and by MURST.} } 
\vskip 0.8cm
\centerline{\bf P. Di Vecchia$^a$\footnote{e-mail:
divecchia@nbivms.nbi.dk}, M. Frau$^{b}$, A. Lerda$^{c,b}$ and A.
Liccardo$^d$}
\vskip .5cm
\centerline{\sl $^a$ NORDITA, Blegdamsvej 17, DK-2100 Copenhagen \O, Denmark}
\vskip .2cm
\centerline{\sl $^b$ Dipartimento di Fisica Teorica, Universit\`a di
Torino} 
\centerline{\sl and I.N.F.N., Sezione di Torino, Via P. Giuria 1, I-10125 
Torino, Italy}
\vskip .2cm
\centerline{\sl $^c$ Dipartimento di Scienze e Tecnologie Avanzate}
\centerline{\sl Universit\`a del Piemonte Orientale, I-15100 
Alessandria, Italy} 
\vskip .2cm
\centerline{\sl $^d$ Dipartimento di Fisica, Universit\`a di Napoli}
\centerline{\sl and I.N.F.N., Sezione di Napoli, 
Mostra d'Oltremare Pad. 19, I-80125 Napoli, Italy}
\vskip 1.3cm
\begin{abstract}
We use the boundary state formalism to provide the full
conformal description of (F,D$p$) bound states. These
are BPS configurations that arise from a superposition of a
fundamental string and a D$p$ brane, and are charged under
both the NS-NS antisymmetric tensor and the $(p+1)$-form
R-R potential. We construct the boundary state for these bound 
states by switching on a constant electric field on the world-volume
of a D$p$ brane and fix its value by imposing the Dirac quantization
condition on the charges. Using the operator formalism we also
derive the Dirac-Born-Infeld action and the classical supergravity
solutions corresponding to these configurations.
\end{abstract}
\end{titlepage}
\newpage
\renewcommand{\thefootnote}{\arabic{footnote}}
\setcounter{footnote}{0}
\setcounter{page}{1}

\sect{Introduction}
\label{introduction}
\vskip 0.5cm
The boundary state, originally introduced for factorizing the planar and 
non-planar open string one-loop diagrams in the closed string 
channel~\cite{EARLY}, has been lately
very useful for describing the D branes \cite{POLC} in the framework
of string theories~\footnote{For references on the use of the boundary
state to study the D branes and their interactions 
see for example Refs.~\cite{BILLO,06}.}. This is because the
boundary state encodes all
relevant properties of the D branes; in fact, as shown in 
Ref.~\cite{cpb}, it reproduces the
couplings of the D branes with the massless closed string states as 
dictated by the Dirac-Born-Infeld action, and also
generates the large distance behavior of the classical 
D brane solutions of supergravity. 

The boundary states which are usually considered in the literature 
describe simple D branes, namely extended
objects that are charged under only one R-R potential.
However, one expects that also more general configurations admit a stringy
description by means of boundary states with a richer structure. 
For example, in Ref.~\cite{cpb} it has already been shown that a
boundary state with an external 
magnetic field describes a D$p$-D$(p-2)$ bound state of two D branes. 
In this paper, instead, we consider boundary states with an 
external electric field and show that
they provide the complete conformal description
of the bound states between a fundamental string
and a D$p$ brane
denoted by (F,D$p$)~\cite{WITTEN}-\cite{HASHI}.
These bound states, which are a generalization of the
dyonic strings introduced in Ref.~\cite{SCHWARZ}, are $p$-dimensional 
extended objects which are charged under both the NS-NS
two-form potential and the R-R $(p+1)$-form potential of the Type II
theories.
Because of this property,
they behave at the same time both as D$p$ branes and 
as fundamental strings. Their nature of D branes allows us to
represent them by means of 
boundary states, while their being also fundamental strings shows up
through the presence of an external electric field on their world-volume.

To describe the (F,D$p$) bound states we introduce a boundary
state containing two parameters: an overall constant $x$ and the constant 
value of the electric field $f$ which can always be taken to be
along one longitudinal direction only. Then, by requiring
the validity of the Dirac quantization condition, we show that these two
parameters can be uniquely fixed in terms of a pair of integers $m$ and $n$
representing, respectively, the charges of the NS-NS antisymmetric tensor
and of the R-R $(p+1)$-form potential.
By projecting this boundary state onto the massless states of the
closed string spectrum \cite{cpb}, we can obtain the long distance 
behavior
of the massless fields that characterize this configuration; 
then we can infer the
complete classical solution describing the
(F,D$p$) bound states and find agreement with the results
recently obtained in
Ref.~\cite{LU2}, which for  $p=1$ reduce to the dyonic strings
of Ref.~\cite{SCHWARZ}. Using the boundary state formalism, 
we also compute the interaction energy between two
(F,D$p$) bound states, and check that the no force condition holds at
the full string level. This fact confirms that the boundary state with an
external electric field provides the complete stringy description of the
BPS bound states formed by fundamental strings and D$p$ branes. 

This paper also contains a
discussion of the limits in which one of the two charges vanishes.
While the limit $m\to 0$, corresponding to a vanishing electric
field, is perfectly under control because in this case the (F,D$p$) bound state
reduces just to a D$p$ brane, the other limit $n\to 0$ is more subtle
because in this case the (F,D$p$) bound state reduces to the 
fundamental string and one does not expect 
that the latter admits a boundary state description. Actually,
when $n\to 0$ the boundary state is not well defined since it contains
an overall vanishing prefactor and a divergent exponential factor involving 
harmonic oscillators. However, when we project it onto the
massless closed string states, 
the vanishing and the divergent factors cancel each other and one is
left with a finite and well-defined expression that exactly reproduces the 
large distance 
behavior of the fundamental string solution. 
Motivated by this observation, we propose a modified 
form of boundary state that generates the fundamental
string solution in the same way as the standard boundary state does for
the D branes. This operator can be
regarded as an {\it effective} conformal description of the fundamental string, 
which, however, cannot be the complete one.

This paper is organized as follows. In \secn{boundary} we write the
boundary state with an external constant gauge field using the
formalism developed in Ref.~\cite{BILLO}. Then, by projecting it along the
massless states of the closed string spectrum, we derive the Dirac-Born-Infeld
action with its Wess-Zumino term. In \secn{fdp} we consider the
boundary state for a D$p$ brane with a constant external
electric field, fix its form by imposing the Dirac quantization
condition on the charges of the NS-NS and R-R potentials, and then obtain the
corresponding classical solutions describing the 
(F,D$p$) bound states. 
In \secn{interaction} we use the boundary state
to compute the interaction energy between two (F,D$p$) bound states and
show the validity of the no-force condition at the string level.
In \secn{Schwarz} we consider in more detail the case $p=1$
and, after introducing non vanishing asymptotic values for 
the scalar fields of the Type IIB theory, show that our boundary 
state reproduces the dyonic string solutions. Finally,
in \secn{conclusions} we discuss the limit $n \to 0$ in which the 
bound state
(F,D$p$) reduces to a fundamental string.
We conclude our paper with two appendices containing more technical details. 
In Appendix A, we introduce the
projectors along the massless closed string states and derive 
some of the formulas used in this paper, while in Appendix B by
performing a T-duality transformation we obtain the boundary state 
corresponding to the bound states (W,D$p$) 
between Kaluza-Klein waves and D$p$ branes
recently discussed in Ref.~\cite{LU2}.

\vskip 1.3cm
\sect{The boundary state with an external field and the D-brane effective
action}
\label{boundary}
\vskip 0.5cm

In this section we are going to briefly review
the construction of the boundary state for a D-brane
with an external field $F$ on its
world-volume. We then show how to use this boundary state
to derive the D-brane low-energy effective action.

\vskip 0.5cm
\subsection{The boundary state with an external field}

In the closed string operator formalism
the supersymmetric D$p$ branes of Type II theories
are described by means of boundary states $\ket{B}$ 
\cite{CALLAN,PCAI}.
These are closed string states which insert a boundary on the
world-sheet and enforce on it the appropriate boundary conditions.
Both in the NS-NS 
and in the R-R sectors, there are two possible 
implementations for the boundary conditions of a D$p$ brane which
correspond to two boundary states $\ket{B, \eta}$, with $\eta=\pm 1$.
However, only the combinations
\begin{equation}
\ket{B}_{\rm NS} = \frac{1}{2} \Big[ \ket{B,+}_{\rm NS} 
- \ket{B,-}_{\rm NS} \Big]
\label{gsons}
\end{equation}
and
\begin{equation}
\ket{B}_{\rm R} = \frac{1}{2} \Big[ \ket{B,+}_{\rm R} + 
\ket{B,-}_{\rm R} \Big]
\label{gsor}
\end{equation}
are selected by the GSO projection in the NS-NS and in the R-R sectors
respectively.
As discussed in Ref.~\cite{BILLO}, the boundary 
state $\ket{B,\eta}$ is the product of a matter part 
and a ghost part
\begin{equation}
\label{bs001}
\ket{B, \eta } = \frac{T_p}{2}\,\ket{B_{\rm mat}, \eta } \ket{B_{\rm 
g}, \eta}~~, 
\end{equation}
where
\begin{equation}
\ket{B_{\rm mat}, \eta} = \ket{B_X} 
\ket{B_{\psi}, \eta}~~~,~~~
\ket{B_{\rm g}, \eta} = \ket{B_{\rm gh}} \ket{B_{\rm sgh}, \eta}~~.
\label{bs000}
\end{equation}
The overall normalization $T_p$ can be unambiguously fixed from
the factorization of amplitudes of closed strings emitted
from a disk \cite{FRAU,cpb} and is the brane tension \cite{lectPOL}
\begin{equation}
T_p= \sqrt{\pi}\left(2\pi\sqrt{\alpha'}\right)^{3-p}~~.
\label{tension}
\end{equation}
The explicit expressions of the various components of 
$\ket{B}$ have been given in Ref.~\cite{BILLO} in the case of a static 
D-brane without any external field on its world-volume.
However, the operator structure of the boundary state does not change 
even when more general configurations are considered and is always of 
the form
\begin{equation} 
\label{bs100} \ket{B_X} = \exp\biggl[-\sum_{n=1}^\infty \frac{1}{n}\,
\a_{-n}\cdot S\cdot
\tilde \a_{-n} \biggr]\,\ket{B_X}^{(0)}~~,
\end{equation}
and
\begin{equation}
\label{bs101}
\ket{B_\psi,\eta}_{\rm NS} = -\ii\, 
\exp\biggl[\ii\,\eta\sum_{m=1/2}^\infty
\psi_{-m}\cdot S\cdot\tilde \psi_{-m}\biggr]
\,\ket{0}
\end{equation}
for the NS-NS sector, and
\begin{equation}
\label{bs102}
\ket{B_\psi,\eta}_{\rm R} = - \exp\biggl[\ii\,\eta\sum_{m=1}^\infty
\psi_{-m} \cdot S\cdot\tilde \psi_{-m}\biggr]
\,\ket{B,\eta}_{\rm R}^{(0)}
\end{equation}
for the R-R sector~\footnote{The unusual phases introduced in 
Eqs. (\ref{bs101}) and (\ref{bs102}) will turn out to be convenient
to study the couplings of the massless closed string states
with a D-brane and to find the correspondence with the
classical D-brane solutions obtained from supergravity. Note that
these phases are instead irrelevant when one computes the interactions
between two D-branes.}.
The matrix $S$ and the zero-mode contributions 
$\ket{B_X}^{(0)}$ and $\ket{B,\eta}_{\rm R}^{(0)}$ 
encode all information about the overlap equations that the
string coordinates have to satisfy, which in turn depend on the 
boundary conditions of the open strings ending on the D$p$ brane.
Since the ghost and superghost fields are not affected by the type of
boundary conditions that are imposed, the ghost part 
of the boundary state is always the same. Its explicit expression 
can be found in Ref.~\cite{BILLO}. We do not write it again here 
since it will not play any significant role for our present purposes.
However, we would like to recall that the
boundary state must be written in the $(-1,-1)$ superghost picture in 
the NS-NS sector, and in the asymmetric $(-1/2,-3/2)$
picture in the R-R in order to saturate the superghost number 
anomaly of the disk \cite{FMS,BILLO}.

When a constant gauge field $F$ is present on the 
D-brane world-volume, the overlap conditions that the boundary
state must satisfy are \cite{CALLAN} 
\bea
&&\left\{ (\one + {\hat{F}})^{\alpha}_{~\beta} \,\alpha_{n}^{\beta} +
(\one - {\hat{F}})^{\alpha}_{~\beta} \,{\tilde{\alpha}}_{-n}^{\beta}
\right\}\ket{B_X }=0
\nonumber \\
&&~~~~~~~(q^i - y^i ) \ket{B_X} = 
\left\{\alpha^i_n -{\tilde{\alpha}_{-n}^i} \right\}\ket{B_X} =0~~~~n \neq 0
\label{bosove1}
\ena
for the bosonic part, and 
\bea
&&\left\{ (\one + {\hat{F}})^{\alpha}_{~\beta} \,\psi_{m}^{\beta} - \ii 
\,\eta \,(\one - {\hat{F}})^{\alpha}_{~\beta} \,
{\tilde{\psi}}_{-m}^{\beta} \right\}\ket{B_{\psi} , \eta } =0
\nonumber \\
&&~~~~~~~~~~~~~~\left\{  \psi_m^i +\ii \,\eta\,{\tilde 
\psi_{-m}^i}\right\} \ket{B_\psi,\eta} =0
\label{bosove2} 
\ena
for the fermionic part. In these equations, the Greek indices 
$\alpha,\beta,\ldots$ label the world-volume directions $0,1,\ldots,p$
along which the D$p$ brane extends, 
while the Latin indices $i,j,\ldots$ label the transverse directions 
$p+1,\ldots,9$; moreover ${\hat{F}} = 2 \pi \alpha F$.
As already noticed in Ref.~\cite{CALLAN}, these equations are solved by 
the ``coherent states'' (\ref{bs100})-(\ref{bs102}) with a matrix 
$S$ given by 
\begin{equation}
{S}_{\mu\nu} = \Big(\big[(\eta - {\hat{F}})(\eta + 
{\hat{F}})^{-1}\big]_{\alpha\beta} \,;\, - \delta_{ij}\Big)
\label{modi4}
\eeq
and with the zero-mode parts given by
\begin{equation}
\ket{B_X}^{(0)} =  \sqrt{-\det (\eta + {\hat{F}})}~
\delta^{(9-p)}(q^i-y^i)\,\prod_{\mu=0}^9\ket{k^\mu=0}
\label{x0f}
\end{equation}
for the bosonic sector, and by
\beq
\label{bsr0}
\ket{B_\psi,\eta}_\R^{(0)} =
\left( C\Gamma^0\Gamma^{1}\ldots
\Gamma^{p} \,
\frac{1+\ii\eta\Gamma_{11}}{1+\ii\eta}\, U\right)_{AB}~~
\,\ket{A} \,\ket{\tilde B}~~
\eeq
for the R sector. In writing these formulas we have denoted by
$y^i$ the position of the D-brane, by $C$ the charge conjugation 
matrix and by $U$ the following matrix
\beq
U = \frac{1}{\sqrt{- \det (\eta + {\hat{F}})}}~
;{\exp}\Big(-\shalf \,{\hat{F}}_{\alpha \beta} \Gamma^{\alpha} 
\Gamma^{\beta}\Big) ;
\label{umat}
\eeq
where the symbol $; \hspace{.5cm} ;$ means that one has to expand the 
exponential and then antisymmetrize the indices of the $\Gamma$-matrices.
Finally, $\ket{A}\,\ket{\tilde B}$ stands for the spinor vacuum of the 
R-R sector \footnote{For our conventions on $\Gamma$-matrices, spinors 
etc. see for example Refs.~\cite{cpb,BILLO}.}.

We would like to remark that the overlap equations (\ref{bosove1}) and 
(\ref{bosove2}) do not allow to determine the overall normalization of 
the boundary state, 
and not even to get the Born-Infeld prefactor of \eq{x0f}. 
The latter was derived in Ref.~\cite{CALLAN}. It can also more easily be
obtained by boosting the boundary state and then performing a 
T-duality as explicitly shown in Ref.~\cite{CANGEMI}. Notice that this
prefactor is present only in the NS-NS component of the boundary state
because in the R-R sector it cancels out if we use the explicit
expression for the matrix $U$ given in \eq{umat}. 

We end this subsection with a few comments. If $F$ is an external 
{\it magnetic} field, the corresponding boundary state describes
a stable BPS bound state formed by a D$p$ brane with other lower 
dimensional D-branes (like for example the D$p$-D$(p-2)$ bound state).
This case was explicitly considered in Ref.~\cite{cpb} where
the long distance behavior of the massless fields of these
configurations was determined
using the boundary state approach.
On the contrary, if $F$ is an external {\it electric} field, then the
boundary state describes a stable bound state between a
fundamental string and a D$p$ brane that preserves one half
of the space-time supersymmetries \cite{WITTEN,SCHMID,SHEIK}. This kind
of bound state denoted by (F,D$p$) is a generalization of the 
dyonic string configurations of Schwarz \cite{SCHWARZ}
which 
has recently been studied from the supergravity
point of view \cite{LU1,LU2} and will be analyzed in detail
in the following sections. However,
before doing this, for completeness we show
how the low-energy effective action of a D-brane is related
to the boundary state we have just constructed.

\vskip 0.5cm
\subsection{The D-brane effective action}
\label{borninfeld}
As we have mentioned before, the boundary state is the 
exact conformal description of a D-brane and therefore it contains
the complete information about the
interactions between a D-brane and the closed strings that propagate in the bulk.
In particular it encodes
the couplings with the bulk massless fields which can be simply obtained by
saturating the boundary state $\ket{B}$ with the massless states of
the closed string spectrum.
In order to find a non-vanishing result, it is necessary to soak up
the superghost number anomaly of the disk and thus, as a consequence of the
superghost charge of the boundary state, we have to use closed string
states in the $(-1,-1)$ picture in the NS-NS sector and states in the asymmetric
$(-1/2,-3/2)$ picture in the R-R sector.

In the NS-NS sector, the states that represent
the graviton $h_{\mu\nu}$, the dilaton $\phi$ and the Kalb-Ramond 
antisymmetric tensor $A_{\mu\nu}$ are of the form
\begin{equation}
\epsilon_{\mu\nu}\,{\tilde\psi}_{-\frac{1}{2}}^\mu\,{\psi}_{-\frac{1}{2}}^\nu
\,\ket{k/2}_{-1}~\ket{\widetilde{k/2}}_{-1}
\label{states}
\end{equation}
with
\begin{equation}
\epsilon_{\mu\nu} = h_{\mu\nu}~~,~~h_{\mu\nu}=h_{\nu\mu}~~,~~
k^\mu h_{\mu\nu}=\eta^{\mu\nu}h_{\mu\nu}=0
\label{gravit}
\end{equation}
for the graviton,
\begin{equation}
\epsilon_{\mu\nu} = \frac{\phi}{2\sqrt{2}}\,(\eta_{\mu\nu} -k_\mu\ell_\nu -
k_\nu\ell_\mu)~~,~~\ell^2=0~~,~~k\cdot\ell=1
\label{dilat}
\end{equation}
for the dilaton, and
\begin{equation}
\epsilon_{\mu\nu} = \frac{1}{\sqrt{2}}\,A_{\mu\nu}~~,~~
A_{\mu\nu}=-A_{\nu\mu}~~,~~k^\mu A_{\mu\nu}=0
\label{kalb}
\end{equation}
for the Kalb-Ramond field \footnote{The factor of $1/\sqrt{2}$ in
\eq{kalb} is necessary to have a canonical normalization, see also
Ref.~\cite{CRAPS}.}. In order to obtain their couplings with the boundary state
it is useful to first compute the quantity
\beq
J^{\mu\nu} \equiv \,{}_{-1}\bra {{\widetilde{k/2}}}~{}_{-1}\bra{k/2}
\,\psi^{\nu}_{\,\frac{1}{2}} \,{\tilde{\psi}}^{\mu}_{\,\frac{1}{2}} 
\, \ket{B}_{\rm NS}=
 - \frac{T_p}{2}\, V_{p+1}\,\sqrt{- \det(\eta + {\hat{F}})} ~ S^{\nu\mu}
\label{coubs}
\eeq
where $V_{p+1}$ is the (infinite) world-volume of the brane,
and then to project it on the various independent fields using their
explicit polarizations. We thus obtain: for the graviton
\beq
J_h \equiv J^{\mu \nu} \,h_{\mu\nu} =- T_p \,V_{p+1}\, \sqrt{- \det 
(\eta + {\hat{F}})} 
~
\Big[(\eta + {\hat{F}})^{-1}\Big]^{\alpha \beta} h_{\beta\alpha}
\label{agra}
\eeq
where we have used the tracelessness of $h_{\mu\nu}$;
for the dilaton
\begin{eqnarray}
J_{\phi} &\equiv & \frac{1}{2\sqrt{2}}\,J^{\mu\nu}\, 
\left( \eta_{\mu\nu}-k_{\mu} \ell_{\nu} - 
k_{\nu}\ell_{\mu}\right) \,\phi
\nonumber \\
&=&\frac{T_p}{2\sqrt{2}}\, V_{p+1} \,\sqrt{- \det 
(\eta + {\hat{F}})} ~
\left[ 3 -p + {\rm Tr} \left( {\hat{F}} (\eta + {\hat{F}})^{-1} \right) \right]
\,\phi~~;
\label{adil}
\end{eqnarray}
and finally for the Kalb-Ramond field
\begin{eqnarray}
J_{A} &\equiv & \frac{1}{\sqrt{2}}\, J^{\mu \nu} \,A_{\mu\nu} =
-\frac{T_p}{2\sqrt{2}} \,V_{p+1} \,\sqrt{- \det (\eta + {\hat{F}})} 
~\Big[(\eta-{\hat{F}})(\eta + {\hat{F}})^{-1} \Big]^{\alpha \beta}A_{\beta \alpha}
\nonumber \\
&=&
-\frac{T_p}{\sqrt{2}}\,V_{p+1}\,\sqrt{- \det (\eta + {\hat{F}})}
~\Big[(\eta + {\hat{F}})^{-1} \Big]^{\alpha \beta}A_{\beta \alpha}
\label{aas}
\end{eqnarray}
where in the second line we have used the antisymmetry of $A_{\mu\nu}$.

We now show that the couplings $J_h$, $J_\phi$ and $J_A$ 
are precisely the ones that are
produced by the Dirac-Born-Infeld action which governs the low-energy
dynamics of the D-brane. In the string frame, this action reads
as follows
\beq
S_{DBI} = - \frac{T_{p}}{\kappa} \int_{V_{p+1}}
 d^{p+1} \xi ~{\rm e}^{- \varphi}
\sqrt{- \det\left[ G + {\cal A} + {\hat F} \right] }
\label{borninfe}
\eeq
where $2 \kappa^{2} = (2\pi)^7 (\alpha ')^4 g_s^2$ is Newton's
constant ($g_s$ being the string coupling), and $G_{\alpha 
\beta}$ and ${\cal
A}_{\alpha \beta}$ are respectively the pullbacks of the space-time
metric and of the NS-NS antisymmetric tensor on the D-brane world volume.

In order to compare the couplings described by this action with the ones 
obtained from the boundary state, it is first necessary to rewrite $S_{DBI}$
in the Einstein frame. In fact, like any string amplitude
computed with the operator formalism, also the couplings $J_h$, $J_\phi$
and $J_A$ are written in the Einstein frame (this property has
been overlooked in the qualitative analysis of Ref.~\cite{SCHMID}).
Furthermore, it is also convenient to introduce canonically normalized fields. 
These two goals can be realized by means of the following field redefinitions
\beq
G_{\mu \nu} = {\rm e}^{\varphi/2} \,g_{\mu \nu}~~~,
~~~ \varphi = \sqrt{2}\,\kappa\,\phi~~~,~~~ {\cal {A}}_{\mu\nu}  = 
\sqrt{2}\,\kappa\,{\rm e}^{\varphi/2}\,A_{\mu\nu}~~.
\label{ein}
\eeq
Using the new fields in \eq{borninfe}, we easily get 
\begin{equation}
S_{DBI} = - \frac{T_{p}}{\kappa} \int_{V_{p+1}} d^{p+1} \xi~
{\rm e}^{-\frac{\kappa\,(3-p)}{2\sqrt{2}}\,\phi}
\sqrt{- \det\left[ \, g + \sqrt{2}\,\kappa\,A
+ {\hat F}\,{\rm e}^{-\frac{\kappa}{\sqrt{2}}\,\phi} \,\right] }~~.
\label{borninfe2}
\end{equation}
By expanding the metric around the flat background 
\beq
g_{\mu \nu} = \eta_{\mu \nu} + 2 \kappa\,h_{\mu \nu}~~,
\label{meexpa}
\eeq
and keeping only the terms which are 
linear in $h$, $\phi$ and $A$, the 
action (\ref{borninfe2}) reduces to the following expression
\begin{eqnarray}
S_{DBI}&\simeq& - ~T_p \int_{V_{p+1}} d^{p+1} \xi \sqrt{- \det \left[ \eta + 
{\hat F}\right] }
~\Bigg\{
\Big[(\eta + {\hat{F}})^{-1}\Big]^{\alpha \beta} h_{\beta\alpha}
\label{expa}
\\
&-&\frac{1}{2\sqrt{2}}\,
\left[ 3 -p + {\rm Tr} \left( {\hat{F}} (\eta + {\hat{F}})^{-1} \right) \right]
\,\phi
+ \frac{1}{\sqrt{2}} \Big[(\eta + {\hat{F}})^{-1} 
\Big]^{\alpha \beta}A_{\beta \alpha}
\Bigg\}~~.
\nonumber
\end{eqnarray}
It is now easy to see that the couplings with the graviton, the dilaton and
the Kalb-Ramond field that can be obtained from this action are
exactly the same as those obtained from the boundary state
and given in Eqs.~(\ref{agra}), (\ref{adil}) and (\ref{aas})
respectively. 

Let us now turn to the R-R sector. As we mentioned above, in this
sector we have to use states in the asymmetric $(-1/2,-3/2)$ 
picture in order to soak up the superghost number anomaly of the disk.
In the more familiar symmetric $(-1/2,-1/2)$ picture the massless
states are associated to the field strengths of the R-R potentials. On the
contrary, in the $(-1/2,-3/2)$ picture
the massless states are associated directly to the R-R potentials
which, in form notation, we denote by
\begin{equation}
C_{(n)} = \frac{1}{n!}\,C_{\mu_1\ldots\mu_n}\,dx^{\mu_1}
\,\wedge\ldots\wedge\,dx^{\mu_n}
\label{potn}
\end{equation}
with $n=1,3,5,7,9$ in the Type IIA theory and $n=0,2,4,6,8,10$ in the
Type IIB theory. The string states $\ket{C_{(n)}}$
representing these potentials have a rather
non-trivial structure. In fact, as shown in Ref.~\cite{BILLO},
the natural expression
\begin{equation}
\label{fred6t}
~~~\ket{C_{(n)}} \simeq \frac{1}{n!}
\,C_{\mu_1\ldots\mu_n}\,
\left(C\Gamma^{\mu_1\ldots\mu_n}\frac{1+\Gamma_{11}}{2}\right)_{AB}
\ket{A;k/2}_{-1/2}~\ket{{\widetilde B};\widetilde{k/2}}_{-3/2}~~~
\end{equation}
is BRST
invariant only if the potential is pure gauge. To avoid this restriction,
in general it is necessary to add to \eq{fred6t} a whole series of terms with the
same structure but with different contents of superghost zero-modes.
However, in the present situation there exists a
short-cut that considerably simplifies the analysis.  
In fact, one can use the incomplete states (\ref{fred6t}) 
and ignore the superghosts, 
whose contribution can then be recovered simply by changing 
at the end the 
overall normalizations of the amplitudes~\footnote{Note that this procedure is
not allowed when the odd-spin structure contributes, see Ref.~\cite{BILLO}.}.
Keeping this in mind, the couplings between the 
R-R potentials (\ref{potn})
and the D$p$ brane can therefore be obtained by computing 
the overlap between the states (\ref{fred6t})
and the R-R component of 
the boundary state, namely
\begin{equation}
J_{C_{(n)}} \equiv   \bra{C_{(n)}}B\rangle_{\rm R}~~.
\label{jcn}
\end{equation}
The evaluation of $J_{C_{(n)}}$ is straightforward, even if a bit
lengthy; some details about this calculation are given 
in Appendix A where the
complete
expression for the asymmetric R-R states is used and 
the contribution of the superghosts is 
explicitly taken into account to obtain the correct
normalization. 
The final result is
\begin{equation}
\label{wnplus1}
J_{C_{(n)}}
=
-\frac{T_p}{16\sqrt{2}\,n!}\,V_{p+1}\,
C_{\mu_1\ldots\mu_n}\,
{\rm Tr}
\left(\Gamma^{\mu_n\ldots\mu_1}\Gamma^0\cdots
\Gamma^p\,;{\rm e}^{-\frac{1}{2}{\hat F}_{\alpha\beta}\Gamma^\alpha
\Gamma^\beta};\right)~~.
\nonumber 
\end{equation}
It is easy to realize that the trace in this equation 
is non-vanishing only if $n=p+1-2\ell$, 
where $\ell$ denotes the power of ${\hat F}$ which is produced by
expanding the exponential term. 
Due to the antisymmetrization $;~~;$
prescription, the integer $\ell$ takes only a finite number of values 
up to a maximum $\ell_{\rm max}$ which is $p/2$ for the
Type IIA string and $(p+1)/2$ for the Type IIB string.
The simplest term to compute, corresponding to $\ell=0$, describes
the coupling of the boundary state with a $(p+1)$-form potential
of the R-R sector and is given by
\begin{equation}
J_{C_{(p+1)}} = \frac{\sqrt{2}\,T_p}{(p+1)!}\,
V_{p+1}\,C_{\alpha_0\ldots\alpha_p}\,\varepsilon^{\alpha_0\ldots\alpha_p}
\label{fite45}
\end{equation}
where $\varepsilon^{\alpha_0\ldots\alpha_p}$ is the completely
antisymmetric tensor on the D-brane world-volume~\footnote{Our convention
is that $\varepsilon^{0\ldots p}=-\varepsilon_{0\ldots p}=1$.}.
{F}rom \eq{fite45} we can immediately
deduce that the charge $\mu_p$ of a D$p$ brane with respect to 
the R-R potential $C_{(p+1)}$ is  
\beq
\mu_p = \sqrt{2} \,T_p
\label{cou89}
\eeq
in agreement with Polchinski's original calculation \cite{POLC}. 

The next term in the expansion of the exponential 
of \eq{wnplus1} corresponds to $\ell=1$ and
yields the coupling of the D$p$ brane with a 
$(p-1)$-form potential which is given by
\begin{equation}
J_{C_{(p-1)}}=
\frac{\mu_p}{2(p-1)!}\,
V_{p+1}\,C_{\alpha_0\ldots\alpha_{p-2}}\,
{\hat F}_{\alpha_{p-1}\alpha_p}\,
\varepsilon^{\alpha_0\ldots\alpha_p}~~.
\label{sete56}
\end{equation}
By proceeding in the same way, one can easily evaluate also the
higher order terms generated by the exponential 
which describe the interactions of the D-brane
with potential forms of lower degree. 
All these couplings can be encoded in the following
Wess-Zumino-like term   
\begin{equation}
S_{\rm WZ} = \mu_p\,\int_{V_{p+1}}\left[
\sum_{\ell=0}^{\ell_{\rm max}}
C_{(p+1-2\ell)}\,\wedge\,
{\rm e}^{\hat F}\right]_{p+1}
\label{WZte}
\end{equation}
where ${\hat F} = \frac{1}{2} \,{\hat F}_{\a \b}\, 
d\xi^{\a} \wedge d\xi^{\b}~$,
and $C_{(n)}$ is the pullback of the $n$-form potential
(\ref{potn}) on the D-brane world-volume. The square bracket
in \eq{WZte} means that in expanding the exponential form
one has to pick up only the terms of total degree $(p+1)$,
which are then integrated over the $(p+1)$-dimensional world-volume.

In conclusion we have explicitly shown that by projecting
the boundary state $\ket{B}$ with an external field
onto the massless states of the closed string spectrum, one can
reconstruct the linear part of the
low-energy effective action of a D$p$ brane. This is the
sum of the Dirac-Born-Infeld part (\ref{expa})
and the (anomalous) Wess-Zumino term (\ref{WZte}) which are produced
respectively by the NS-NS and the R-R components of the boundary state.

\vskip 1.3cm
\sect{(F,D$p$) bound states from the boundary state}
\label{fdp}
\vskip 0.5cm
In this section we are going to show that 
the boundary state
constructed before can be used to obtain the long 
distance behavior of the various fields
describing the bound state (F,D$p$) formed by a 
fundamental string and a D$p$ brane. This type of bound
state is a generalization of the dyonic string solution of Schwarz
\cite{SCHWARZ} and
has been recently discussed from the supergravity point of view 
\cite{LU1,LU2}.
As we mentioned before, the (F,D$p$) 
bound state can be obtained from a D$p$ brane by turning on 
an {\it electric} field ${\hat F}$ on its world
volume \cite{WITTEN,SCHMID,SHEIK}.
With no loss in generality we can choose ${\hat F}$ 
to have non vanishing components only in the directions $X^0$ and $X^1$ 
so that it can be represented by the following $(p+1)\times (p+1)$ 
matrix 
\begin{equation}
{\hat F}_{\alpha\beta} =  
\left( \begin{array}{ccccc} 
           0& -f & & &  \\
                       f & 0& & & \\
			 & & 0 & &  \\
			 & & &\ddots &  \\
			 & & & & 0
			 \end{array} \right)~~.
\label{effe3}
\end{equation}
Using this expression in \eq{modi4} one can easily see that the 
longitudinal part of the matrix $S$ appearing in the boundary
state is given by
\begin{equation}
S_{\alpha\beta} =  
\left( \begin{array}{ccccc} 
           -\frac{1 + f^2}{1-f^2}  & \frac{2f}{1-f^2}& & &  \\
                       -\frac{2f}{1-f^2} & \frac{1+f^2}{1-f^2} & & & \\
			 & & 1 & &  \\
			 & & &\ddots &  \\
			 & & & & 1
			 \end{array} \right)
\label{emme3}
\end{equation}
while the transverse part of  $S$ is simply minus the identity
in the remaining $(9-p)$ directions. 
Furthermore, using \eq{effe3} one finds
\begin{equation}
- \det \left(\eta + {\hat{F}} \right) = 1 - f^2~~.
\label{emme31}
\end{equation}
As we have discussed in Ref.~\cite{cpb}, the boundary state can be 
used in a very efficient way to obtain the long distance behavior of 
the fields emitted by a D-brane and obtain the corresponding
classical solution at long distances. To do so one simply
adds a closed string propagator $D$ to the boundary state $B$ and then
projects the resulting expression onto the various 
massless states of the closed string spectrum. \
According to this procedure, the 
long-distance fluctuation of a field $\Psi$ is then given by 
\begin{equation}
\delta \Psi \equiv \bra{P^{(\Psi)}} D \ket{B}
\label{ml1}
\end{equation}
where $\bra{P^{(\Psi)}}$ denotes the projector associated to $\Psi$. 
The explicit expressions for these projectors can be found in
Appendix A (see Eqs.(\ref{dila45}), (\ref{gravi87}), (\ref{anti76}), 
(\ref{sta65})) for all massless fields of the NS-NS and R-R sectors.

Before giving the details of this calculation, we would like
to make a few comments. Firstly, since we are not using
explicitly the ghost and superghost degrees of freedom, we must
take into account their contribution by shifting appropriately
the zero-point energy and use for the closed string
propagator the following expression
\begin{equation}
D = \frac{\alpha '}{4 \pi} \int_{|z| \leq 1} \frac{d^2 z}{|z|^2}\, 
z^{L_0 - a} \,{\bar{z}}^{{\tilde{L}}_0 - a}
\label{propa}
\end{equation}
where the operators $L_0$ and ${\tilde{L}}_0$ depend only on the 
orbital oscillators and the intercept is $a=1/2$ in the NS-NS sector 
and $a=0$ in the R-R sector.  Secondly, since we want to describe
configurations of branes with arbitrary R-R charge, we multiply the
entire boundary state by an overall factor of $x$. Later we 
will see that the consistency of the entire construction will require 
that $x$ be an integer, and also that the electric field strength $f$
cannot be arbitrary.

Let us now begin our analysis by studying the projection
(\ref{ml1}) in the NS-NS sector. Since all 
projectors onto the NS-NS massless fields contain 
the following structure
\begin{equation}
\label{projal}
{}_{-1}\bra {{\widetilde{k/2}}}~{}_{-1}\bra{k/2}
\,\psi^{\nu}_{\,\frac{1}{2}} \,{\tilde{\psi}}^{\mu}_{\,\frac{1}{2}}
\end{equation}
as we can see from the
explicit expressions given in Eqs. (\ref{dila45}) - (\ref{anti76}),
it is first convenient to compute the matrix element
\begin{equation}
T^{\mu \nu} \equiv\,{}_{-1}\bra {{\widetilde{k/2}}}~{}_{-1}\bra{k/2}
\,\psi^{\nu}_{\,\frac{1}{2}} \,{\tilde{\psi}}^{\mu}_{\,\frac{1}{2}} |\, 
D \, \ket{B}_{\rm NS}
= -x \,\frac{T_p}{2} \,\frac{V_{p+1}}{k_{\bot}^{2}}
 \,\sqrt{1-f^2} ~ S^{\nu\mu}
\label{flu34}
\end{equation}
where $k_\bot$ is the momentum in the transverse directions which is
emitted by the brane. Notice that the matrix 
$T^{\mu\nu}$ differs from
the matrix $J^{\mu\nu}$ defined in \eq{coubs} simply by the factor of
$1/k_{\bot}^{2}$ coming from the
insertion of the propagator, and by the overall normalization 
({\it i.e.} the factor of $x$).

Using this result and the explicit form of the dilaton projector 
(\ref{dila45}),
after some straightforward algebra, we find that the long-distance
behavior of the dilaton of the (F,D$p$) bound state is given by
\begin{equation}
\delta \phi  \equiv \bra{P^{(\phi)}} D \ket{B}_{\rm NS}
=\frac{1}{2\sqrt{2}}\,\left(\eta^{\mu\nu} -
k^{\mu} \ell^{\nu} - k^{\nu}\ell^{\mu}\right)
T_{\mu\nu}~~.
\label{dil45}
\end{equation}
Using the explicit expression for the matrix $T_{\mu\nu}$ we get
\begin{equation}
\label{dil4510}
\delta \phi = \mu_p \, \frac{V_{p+1} }{k_{\bot}^{2}}\,
x \frac{ f^2(p -5)+ (3-p) }{4\,\sqrt{1 - f^2}}
\end{equation}
where $\mu_p$ is the unit of R-R charge of a D$p$ brane defined
in \eq{cou89}.
Similarly, using the projector (\ref{anti76}) for the antisymmetric
Kalb-Ramond field, we find
\begin{equation}
\delta A_{\mu\nu} \equiv \bra{P_{\mu\nu}^{(A)}} D \ket{B}_{\rm NS} =
\frac{1}{\sqrt 2}\Big( T_{\mu\nu} - T_{\nu\mu} \Big)~~.
\label{anti67}
\end{equation}
Since with our choices the matrix $T_{\mu\nu}$ is symmetric except in 
the block of the $0$ and $1$ directions (see \eq{emme3}), we immediately 
conclude that the only non-vanishing component of the Kalb-Ramond field
emitted by the (F,D$p$) bound state is  $A_{01}$ whose long-distance
behavior is given by
\begin{equation} 
\delta A_{01}= {\mu_p}\,\frac{V_{p+1} }{k_{\bot}^{2}}\,
\frac{ x\,f }{\sqrt{1 - f^2}}~~.
\label{anti671}
\end{equation}
Finally, using \eq{gravi87} we find that the components of the metric tensor 
are 
\begin{equation}
\delta h_{\mu \nu} \equiv \bra{P_{\mu\nu}^{(h)}} D \ket{B}_{\rm NS} = 
\frac{1}{2}\Big(T_{\mu\nu}+T_{\nu\mu}\Big) 
-\frac{\delta\phi}{2\sqrt{2}}\, \,\eta_{\mu\nu}
\label{gra3}
\end{equation}
which explicitly read
\begin{eqnarray}
\label{gra31}
\delta h_{00} &=& - \delta h_{11} = 
{\mu_p}\,\frac{V_{p+1}}{k_{\bot}^2}\,x\,\frac{ f^2(p-1)+ (7-p)}
{8\sqrt{2}\sqrt{1-f^2}}~~,
\nonumber \\
\delta h_{22} &=& \ldots ~= \delta h_{pp} = 
{\mu_p}\,\frac{V_{p+1}}{k_{\bot}^2}\,x\,\frac{ f^2(9-p)+ (p-7)}
{8\sqrt{2}\sqrt{1-f^2}}~~,
\\
\delta h_{p+1,p+1}&=& \ldots ~= \delta h_{99} =
{\mu_p}\,\frac{V_{p+1}}{k_{\bot}^2}\,x\,\frac{ f^2(1-p)+ (p+1)}
{8\sqrt{2}\sqrt{1-f^2}}~~.
\nonumber 
\end{eqnarray}

Let us now turn to the R-R sector. In this case, after
the insertion of the closed string propagator, we have to 
saturate the R-R boundary state (\ref{gsor}) with the projectors on the
various R-R massless fields given in \eq{sta65}. 
This calculation is completely analogous to the one described 
in the previous section and performed in detail in
Appendix A to obtain the couplings of a D$p$ brane with the R-R 
potentials. The only new features are the 
overall factor of $x$ and the presence of the factor of
$1/k_\bot^2$ produced by the closed string
propagator.
Due to the structure of the R-R component
of the boundary state
describing the bound state (F,D$p$), it is not difficult
to realize that the only projectors of the form (\ref{sta65}) that can give 
a non vanishing result are those corresponding to a $(p+1)$-form and 
to a $(p-1)$-form with all indices along the world-volume
directions. In particular, we find that the long distance
behavior of the $(p+1)$-form is given by
\begin{equation}
\delta C_{01\cdots p} 
\equiv \bra{P^{(C)}_{01\cdots p}} \,D\,\ket{B}_{\rm R} 
= - \mu_p \,\frac{V_{p+1}}{k_{\bot}^2}\,x~~.
\label{antiR2}
\end{equation}

Similarly, given our choice of the
external field, we find that the only non vanishing component 
of the $(p-1)$-form emitted by the boundary state
is $C_{23\cdots p}$ whose long-distance behavior turns out to be
\begin{equation}
\delta C_{23\cdots p} \equiv \bra{P^{(C)}_{23\cdots p}}\,D \,\ket{B}_{\rm R} =
- \mu_p\, \frac{ V_{p+1}}{k_{\bot}^{2}} \,x\,f~~. 
\label{chi78}
\end{equation}
Notice that if $p=1$ this expression has to be interpreted as the
long-distance behavior of the R-R scalar which is usually denoted
by $\chi$.

In all our previous analysis, the two parameters $x$ and $f$
that appear in the boundary state seem to be arbitrary. However,
this is not so at a closer inspection. In fact, 
they are strictly related
to the electric charges of the (F,D$p$) configuration under the Kalb-Ramond field
and the R-R $(p+1)$-form potential~\footnote{Notice that
these charges are indeed of electric
type since the only non-vanishing components of the corresponding
potentials have an index along the time direction.}. 
It is well-known that these charges
must obey the Dirac quantization condition, {\it i.e.} they
must be integer multiples of the fundamental unit of (electric)
charge of a $p$-dimensional extended object $\mu_p$. In our notations
this quantization condition amounts to 
impose that the coefficients of  
$-\mu_p \,\frac{ V_{p+1}}{k_{\bot}^{2}}$ in Eqs. (\ref{anti671}) and
(\ref{antiR2}) be integer numbers (see also 
\secn{Schwarz} for a discussion of this issue)~\footnote{This particular 
choice
of signs is of course just a matter of convention; as we will see it
leads to the results that are usually reported in the literature.}. 
This implies that
\begin{equation}
x = n  ~~~~\mbox{and}~~~~ - \frac{x f}{\sqrt{1 - f^2}} = m
\label{xf}
\end{equation}
with $n$ and $m$ two integers. While the restriction on $x$
had to be expected from the very beginning because $x$ simply represents
the number of D$p$ branes (and hence of boundary states) that
form the bound state, the restriction on
the external field $f$ is less trivial. 
In fact, from \eq{xf} we see that $f$
must be of the following form
\begin{equation}
f = - \frac{m}{\sqrt{n^2+m^2}}~~.
\label{xf3} 
\end{equation}
This is precisely the same expression that appears in the
analysis of Ref.~\cite{SCHMID} on the dyonic string configurations, and is
also consistent with the results of Ref.~\cite{LU1,LU2}. 

Using \eq{xf}, we can now rewrite the long distance
behavior of the massless fields produced by a (F,D$p$)
bound state in a more suggestive way. In doing so, we also
perform a Fourier transformation to work in configuration
space. This is readily computed by observing that,
for $p <7$, one has
\begin{equation}
\int d^{(p+1)}x\,d^{(9-p)}y \frac{{\rm e}^{i k_{\bot} \cdot y}}{(7-p)\,
r^{7-p}\,\Omega_{8-p}}= \frac{V_{p+1}}{k_\bot^2}
\label{ft}
\end{equation}
where $\Omega_q=2\pi^{(q+1)/2}/\Gamma((q+1)/2)$ is the area of a unit
$q$-dimensional sphere and $r^2=y_iy^i$ measures the distance 
from the branes.
For later convenience, we also introduce the following
notations
\begin{equation}
\Delta_{m,n} = {m^2+n^2}
\label{Delta}
\end{equation}
and
\begin{equation}
Q_p = \mu_p\,\frac{\sqrt{2}\,\kappa\,
\Delta_{m,n}^{1/2}}{(7-p)\,\Omega_{8-p}}~~.
\label{Qp}
\end{equation}
Then, using \eq{dil4510} and assuming for the time being that the dilaton has vanishing
vacuum expectation value, after some elementary steps, we obtain that the
long-distance behavior of the dilaton is
\begin{equation}
\varphi = \sqrt{2}\,\kappa\,\phi
\simeq \left(-\frac{1}{2}+\frac{5-p}{4}\,\frac{n^2}{\Delta_{m,n}}
\right)\,\frac{Q_p}{r^{7-p}}~~.
\label{dilft}
\end{equation}
Since we are going to compare our results with the
standard supergravity description of D-branes, we have reintroduced
the field $\varphi$ which differs from the canonically normalized
dilaton $\phi$ by a factor of $\sqrt{2}\,\kappa$ (see also \eq{ein}).
Similarly, recalling that $g_{\mu\nu}=\eta_{\mu\nu}+2\kappa\,h_{\mu\nu}$,
from \eq{gra31} we find 
\begin{eqnarray}
g_{00} & = & -g_{11} \simeq -1 -
\left(-\frac{3}{4}+\frac{p-1}{8}\,\frac{n^2}{\Delta_{m,n}}
\right)\,\frac{Q_p}{r^{7-p}}~~,
\nonumber \\
g_{22}&=&\ldots~=~g_{pp} \simeq
1 + \left(\frac{1}{4}+\frac{p-9}{8}\,\frac{n^2}{\Delta_{m,n}}
\right)\,\frac{Q_p}{r^{7-p}}~~,
\label{gft}
\\
g_{p+1,p+1} & = & \ldots~=~g_{99} 
\simeq 1 + \left(\frac{1}{4}+\frac{p-1}{8}\,\frac{n^2}{\Delta_{m,n}}
\right)\,\frac{Q_p}{r^{7-p}}~~.
\nonumber
\end{eqnarray}
Rescaling the Kalb-Ramond field by a factor of
$\sqrt{2}\,\kappa$ to obtain the standard
supergravity normalization and using \eq{anti671}, we easily
get
\begin{equation}
\label{aft}
{\widehat A} = \sqrt{2}\kappa\,A \simeq 
-\frac{m}{\Delta_{m,n}^{1/2}}\,\frac{Q_p}{r^{7-p}}
~dx^0\wedge dx^1~~.
\end{equation}
Finally, repeating the same steps for the R-R potentials
(\ref{antiR2}) and (\ref{chi78}) we find
\begin{equation}
\label{cpft}
{\widehat C}_{(p+1)} = \sqrt{2}\kappa\,
C_{(p+1)}\simeq 
-\frac{n}{\Delta_{m,n}^{1/2}}\,\frac{Q_p}{r^{7-p}}
~dx^0\wedge \ldots \wedge dx^p~~,
\end{equation}
and
\begin{equation}
\label{cp1ft}
{\widehat C}_{(p-1)} = \sqrt{2}\kappa\,
C_{(p-1)}\simeq 
\frac{m}{n}\,\frac{n^2}{\Delta_{m,n}}\,\frac{Q_p}{r^{7-p}}
~dx^2\wedge \ldots \wedge dx^p~~.
\end{equation}
Eqs. (\ref{dilft})-(\ref{cp1ft}) represent the leading
long-distance behavior of the massless fields emitted
by the (F,D$p$) bound state. 
Of course the simple knowledge of this asymptotic
behavior is not sufficient to determine the exact and
complete form of the corresponding classical brane-solution.
To do this, one would need
to compute also the higher order terms in the large distance expansion,
and eventually resum the series. These higher order terms
correspond to one-point functions for the
massless closed string states emitted 
by world-sheets with many boundaries, and thus
in our formalism one would need to compute one-point amplitudes 
with many boundary states.
Clearly these calculations become more and more
involved as the number of boundary states increases,
and so far only the next-to-leading term in
the case of a single D3 brane has been
computed \cite{LEIGH},
even if with a different formalism.
However, to obtain the complete brane-solution
one can follow an alternative
(and easier) route, namely one can make an
{\it Ansatz} on the form of the solution, use
the leading long-distance behaviour to fix 
the parameters and finally check that the
classical field equations are satisfied.
In our case, it is reasonable to assume that
the exact (F,D$p$) brane solution
can be written entirely in terms of $p$-dependendent
powers of harmonic functions. An inspection of 
Eqs. (\ref{dilft})-(\ref{gft}) suggests to introduce 
two harmonic functions, namely
\begin{equation}
\label{hr}
H(r) = 1 + \frac{Q_p}{r^{7-p}}
\end{equation}
which is the usual function appearing in the D-brane
solutions, and 
\begin{equation}
\label{h1r}
H'(r) = 1 + \frac{n^2}{\Delta_{m,n}}~\frac{Q_p}{r^{7-p}}
\end{equation}
which has been introduced also in Ref.~\cite{LU2}. 
Then, according to our assumptions and using 
Eqs. (\ref{dilft})-(\ref{cp1ft}), we can infer that the
dilaton is
\begin{equation}
{\rm e}^{\varphi} = H^{-{1}/{2}}\,{H'}^{{(5-p)}/{4}}~~,
\label{dilfin}
\end{equation}
the metric is
\begin{eqnarray}
\label{metricfin}
ds^2 &=& H^{-{3}/{4}}\,{H'}^{{(p-1)}/{8}}
\left[-\left(dx^0\right)^2+\left(dx^1\right)^2
\right] \nonumber \\
&+&
H^{{1}/{4}}\,{H'}^{{(p-9)}/{8}}
\left[\left(dx^2\right)^2+\cdots+\left(dx^p\right)^2
\right] \nonumber \\
&+&
H^{{1}/{4}}\,{H'}^{{(p-1)}/{8}}
\left[\left(dx^{p+1}\right)^2+\cdots+\left(dx^9\right)^2
\right] ~~,
\end{eqnarray}
the Kalb-Ramond 2-form is 
\begin{equation}
\label{krfin}
{\widehat A} = \frac{m}{\Delta_{m,n}^{1/2}}~\left({H}^{-1}-1\right)
\,dx^0\wedge dx^1~~,
\end{equation}
and finally the R-R potentials are
\begin{equation}
\label{cpfin}
{\widehat C}_{(p+1)} = \frac{n}{\Delta_{m,n}^{1/2}}~\left({H}^{-1}-1\right)
\,dx^0\wedge \cdots \wedge dx^p~~,
\end{equation}
and
\begin{equation}
\label{cp1fin}
{\widehat C}_{(p-1)} = -\frac{m}{n}~\left({H'}^{-1}-1\right)
\,dx^2\wedge \cdots \wedge dx^p~~.
\end{equation}
Notice that the field strength associated to
${\widehat C}_{(p+1)}$ is electric, whereas
the one associated to ${\widehat C}_{(p-1)}$ is magnetic.
In the case $p=1$, the last equation has to be replaced by
\begin{equation}
\label{axionfin}
\chi = -\frac{m}{n}~\left({H'}^{-1}-1\right)
\end{equation}
where $\chi$ is the R-R scalar field also called axion.

In writing these formulas we have assumed that all fields except the
metric have vanishing asymptotic values. This explains why we have
subtracted the 1 in the last four equations. At this point we 
should check that Eqs. (\ref{dilfin})-(\ref{cp1fin}) are a solution
to the classical field equations. This is indeed the case since
our fields agree with the ones recently derived in Ref.~\cite{LU2} from
the supergravity point of view~\footnote{Actually, in comparing our
results with those of Ref.~\cite{LU2}, we find total agreement
except for the overall sign in the Kalb-Ramond 2-form. Our sign however
agrees with the dyonic string solution of Schwarz \cite{SCHWARZ}
when we put $p=1$.}. Moreover, \eq{axionfin} can be shown to exactly
agree with the axion field of the dyonic string solution of Schwarz
\cite{SCHWARZ} in the case of vanishing asymptotic background values
for the scalars ($\varphi_0 = \chi_0 = 0$).

\vskip 1.3cm
\sect{Interaction between two (F,D$p$) bound states}
\label{interaction}
\vskip 0.5cm

We now analyze some properties of the (F,D$p$) bound
states we have described in the previous section; in particular we
compute the interaction energy between two of them
both from the classical and from the string point of
view. To this aim, let us start by considering the
contribution to the classical interaction energy due
to the exchange of dilatons. The coupling $J_\phi$ of
the dilaton with the boundary state describing the
(F,D$p$) configuration is given in \eq{adil} with an overall
factor of $n$; 
after using Eqs. (\ref{effe3}) and (\ref{xf}), $J_\phi$
explicitly becomes
\begin{equation}
J_\phi = \frac{T_p}{2\sqrt{2}}\,
V_{p+1}\,\frac{n^2\,(3-p)-2m^2}{\Delta_{m,n}^{1/2}}\,\phi~~.
\label{jphi}
\end{equation}
Using this coupling, we can compute the potential
energy density as follows
\begin{equation}
U_\phi = \frac{1}{V_{p+1}}\, 
{\underbrace{J_\phi~J_{\phi}}}
= \frac{T_p^2}{8}\,V_{p+1}\,\frac{\left[
n^2\,(3-p)-2m^2
\right]^2}{\Delta_{m,n}}\,{\underbrace{\phi~\phi}}
\label{uphi}
\end{equation}
where 
\begin{equation}
{\underbrace{\phi~\phi}} = \frac{1}{k_\bot^2}
\label{dilprop}
\end{equation}
is the dilaton propagator. Thus, we have
\begin{equation}
U_\phi = \frac{T_p^2}{8}
\,\frac{\left[
n^2\,(3-p)-2m^2
\right]^2}{\Delta_{m,n}}\,\frac{V_{p+1}}{k_\bot^2}~~.
\label{uphi1}
\end{equation}
Notice that $U_\phi$ is always positive, which implies
that the force between two (F,D$p$) bound states
due to dilaton exchanges is always attractive.

Let us now turn to the contribution to the potential energy
due to graviton exchanges. The coupling of the graviton with
the boundary state is given by \eq{agra} (again with an overall
factor of $n$) which in 
our specific case becomes
\begin{equation}
J_h = - T_p\,V_{p+1}\,\frac{n^2}{\Delta_{m,n}^{1/2}}
\, V^{\alpha\beta}h_{\beta\alpha}
\label{jh}
\end{equation}
where $V^{\alpha\beta}$ is the following
$(p+1)\times (p+1)$ matrix
\begin{equation}
V^{\alpha\beta} = \Big[(\eta + {\hat{F}})^{-1}\Big]^{\alpha \beta} 
=
\left( \begin{array}{ccccc} 
           -\frac{\Delta_{m,n}}{n^2}  & 
	   \frac{m\,\Delta_{m,n}^{1/2}}{n^2}& & &  \\
                       -\frac{m\,\Delta_{m,n}^{1/2}}{n^2}& 
		       \frac{\Delta_{m,n}}{n^2} & & & \\
			 & & 1 & &  \\
			 & & &\ddots &  \\
			 & & & & 1
			 \end{array} \right)
\label{defv}
\end{equation}
and $h_{\beta\alpha}$ is the
graviton polarization. The gravitational potential energy
is then
\begin{equation}
U_h = \frac{1}{V_{p+1}}\, {\underbrace{J_h~J_h}}
= T_p^2\,V_{p+1}\,\frac{n^4}{\Delta_{m,n}}\,
V^{\alpha\beta}\,V^{\gamma\delta}\,
{\underbrace{h_{\beta\alpha}~h_{\delta\gamma}}}
\label{uh}
\end{equation}
where 
\begin{equation}
{\underbrace{h_{\beta\alpha}~h_{\delta\gamma}}}
= \left[\frac{1}{2}\left(\eta_{\beta\delta}
\eta_{\alpha\gamma} + \eta_{\alpha\delta}
\eta_{\beta\gamma}\right) - \frac{1}{8}\,
\eta_{\alpha\beta}\eta_{\gamma\delta}\right]\,
\frac{1}{k_\bot^2}
\label{gravpro}
\end{equation}
is the graviton propagator in the de Donder gauge.
Using the explicit expression of the matrix $V$ given
in \eq{defv} and performing some elementary algebra
we obtain
\begin{equation}
U_h = \frac{T_p^2}{8}
\,\frac{\left[
n^4\,(7-p)\,(p+1)+4n^2\,m^2\,(7-p)+12m^4
\right]}{\Delta_{m,n}}\,\frac{V_{p+1}}{k_\bot^2}~~.
\label{uh1}
\end{equation}
Notice that this gravitational potential energy is always
positive (for $p\leq 7$), signaling the well-known fact that the 
exchange of gravitons always yields an attractive force.

Now let us consider the interaction between two
(F,D$p$) bound states due to exchanges of Kalb-Ramond
antisymmetric tensors. The coupling between the boundary state
and the Kalb-Ramond field is given by \eq{aas} with an overall
factor of $n$ and
in our present case it becomes
\begin{equation}
J_A = - \frac{T_p}{\sqrt{2}}\,V_{p+1}\,
\frac{n^2}{\Delta_{m,n}^{1/2}}\,V^{\alpha\beta}
A_{\beta\alpha}~~.
\label{ja}
\end{equation}
Thus, the corresponding potential energy density is
\begin{equation}
U_A = \frac{1}{V_{p+1}}\,
{\underbrace{J_A~J_A}}
= \frac{T_p^2}{2}\,V_{p+1}\,\frac{n^4}{\Delta_{m,n}}\,
V^{\alpha\beta}\,V^{\gamma\delta}\,
{\underbrace{A_{\beta\alpha}~A_{\delta\gamma}}}
\label{ua}
\end{equation}
where 
\begin{equation}
{\underbrace{A_{\beta\alpha}~A_{\delta\gamma}}}
= \left(\eta_{\beta\delta}
\eta_{\alpha\gamma} - \eta_{\alpha\delta}
\eta_{\beta\gamma}\right)\,
\frac{1}{k_\bot^2}
\label{apro}
\end{equation}
is the propagator of an antisymmetric 2-index tensor in the
Lorentz gauge. Using the explicit form 
(\ref{defv}) of the matrix $V$ and
inserting \eq{apro} into \eq{ua}, we find
\begin{equation}
U_A = - 2\,T_p^2
\,m^2\,\frac{V_{p+1}}{k_\bot^2}~~.
\label{ua1}
\end{equation}
Notice that $U_A$ is always negative meaning that the corresponding
force is always repulsive. This is indeed what should happen
because the (F,D$p$) bound state carries the electric charge
of the Kalb-Ramond field, and two alike charges always
repel each other.

Finally, we compute the interaction energy density
due to the exchange of the R-R potentials. 
In the case of the top (electric) form $C_{(p+1)}$,
the coupling with the boundary state is given by
\eq{fite45} multiplied by $n$ and in our case it explicitly reads
\begin{equation}
J_{C_{(p+1)}} = \sqrt{2}\,T_p\,V_{p+1}\,n\,C_{01\cdots p}~~.
\label{jcp+1}
\end{equation}
Then it is immediate to realize that the potential
energy density is given by
\begin{eqnarray}
U_{C_{(p+1)}} &=& \frac{1}{V_{p+1}}\,
{\underbrace{J_{C_{(p+1)}}~J_{C_{(p+1)}}}}
= 2\,T_p^2\,V_{p+1}\,n^2
\,{\underbrace{C_{01\cdots p}~C_{01\cdots p}}}
\nonumber \\
&=& - 2\,T_p^2\,n^2\,\frac{V_{p+1}}{k_\bot^2}
\label{ucp}
\end{eqnarray}
where we have used the propagator
\begin{equation}
{\underbrace{C_{01\cdots p}~C_{01\cdots p}}} = 
-\frac{1}{k_\bot^2}
\label{cprop}
\end{equation}
which is the obvious generalization of \eq{apro}.
Since $U_{C_{(p+1)}}$ is negative, the corresponding
force is repulsive as it should be, since the (F,D$p$)
bound states carry the same electric charge under
the $(p+1)$-form potential. In a similar way, we can
compute the contribution to the interaction due to the exchange
of the magnetic R-R potentials $C_{(p-1)}$. In this case
the coupling with the boundary state, which we can read from
\eq{sete56} with an overall factor of $n$, is
\begin{equation}
J_{C_{(p-1)}} = \sqrt{2}\,T_p\,V_{p+1}\,\frac{mn}{
\Delta_{m,n}^{1/2}}\,C_{23\cdots p}
\label{jcp-1}
\end{equation}
and hence, the corresponding potential energy 
density turns out to be
\begin{eqnarray}
U_{C_{(p-1)}} &=& \frac{1}{V_{p+1}}\,
{\underbrace{J_{C_{(p-1)}}~J_{C_{(p-1)}}}}
= 2\,T_p^2\,V_{p+1}\,\frac{m^2n^2}{\Delta_{m,n}}
\,{\underbrace{C_{23\cdots p}~C_{23\cdots p}}}
\nonumber \\
&=& 2\,T_p^2\,\frac{m^2n^2}{\Delta_{m,n}}
\,\frac{V_{p+1}}{k_\bot^2}
\label{ucp-1}
\end{eqnarray}
where we have used the propagator
\begin{equation}
{\underbrace{C_{23\cdots p}~C_{23\cdots p}}} = 
\frac{1}{k_\bot^2}~~.
\label{cprop1}
\end{equation}
Notice that this last contribution is positive so that
the associated force is always attractive as it
should be for the exchange of abelian potentials
of magnetic type.
It is interesting to observe that not all 
the NS-NS (or R-R) massless fields contribute with the
same sign to the interaction energy between two
(F,D$p$) bound states. This is to be compared
with what happens with simple D-branes, where the distinction
between attractive and repulsive contributions coincides
with the distinction between the NS-NS and R-R sectors.

In order to compute the mass density, or the tension, of the
(F,D$p$) bound states, we follow Polchinski's
approach \cite{lectPOL}, namely we consider the total
{\it attractive} potential energy and then compare it with Newton's
law in $d=10$ for two $p$-dimensional extended 
objects~\footnote{In the case of simple D$p$ branes this amounts
to consider just the contribution of the NS-NS sector; in our case
however, we cannot use this identification 
since in the NS-NS sector there is a repulsive
contribution and an attractive contribution appears in the R-R sector.}.
If we sum $U_\phi$, $U_h$ and $U_{C_{(p-1)}}$, remarkable simplifications
occur yielding
\begin{equation}
U_{\rm attr} = U_\phi+U_h+U_{C_{(p-1)}} = 2\,T_p^2\,\Delta_{m,n}\,
\frac{V_{p+1}}{k_\bot^2}~~.
\label{uattr}
\end{equation}
Performing a Fourier transformation and using \eq{ft}, we get
\begin{equation}
U_{\rm attr}(r) = \frac{2\,T_p^2\,\Delta_{m,n}}{(7-p)\,\Omega_{8-p}}
\,\frac{1}{r^{7-p}}
\label{uattr1}
\end{equation}
where $r$ is the distance between the two bound states. On the other hand,
Newton's law for two $p$-dimensional extended objects of mass density $M_p$
in $d=10$ reads 
\begin{equation}
U(r) = \frac{2\kappa^2\,M_p^2}{(7-p)\,\Omega_{8-p}}
\,\frac{1}{r^{7-p}}
\label{newton}
\end{equation}
where $2\kappa^2$ is Newton's constant. By comparing Eqs.
(\ref{uattr1}) and (\ref{newton}) we conclude that
\begin{equation}
M_p = \frac{1}{\kappa}\,T_p\,\Delta_{m,n}^{1/2}
\label{massp}
\end{equation}
so that the tension of the (F,D$p$) bound state is
\begin{equation}
T(m,n) = T_p\,\Delta_{m,n}^{1/2} = T_p\,\sqrt{m^2+n^2}~~.
\label{tmn}
\end{equation}
This formula agrees with the one obtained in Refs.~\cite{LU1,LU2} with
ADM considerations, and explicitly shows that the (F,D$p$) configuration is
a non-threshold bound state between a fundamental string of charge $m$
and a D$p$ brane of charge $n$ (in units of $\mu_p$). It
also makes evident the fact that elementary bound states are realized
if $m$ and $n$ are relative prime integers.

If we now compute the total energy density, we find a vanishing result,
{\it i.e.}
\begin{equation}
U_{\rm tot} = U_\phi +U_h +U_A +U_{C_{(p+1)}}+ U_{C_{(p-1)}} = 0
\label{utot}
\end{equation}
meaning that the (F,D$p$) bound states are BPS configurations
satisfying the no-force condition. Actually, this property can be
proved at the full string level by computing the vacuum amplitude
between two boundary states at a distance $r$ from each other, which is
defined by 
\begin{equation}
\Gamma = \bra{B}\,D\,\ket{B}
\label{vacampl}
\end{equation}
where $D$ is the closed string propagator (\ref{propa}). 
(The explicit form of the
conjugate boundary state that has to be used in \eq{vacampl}
can be found at the end of Appendix A.)
Then following the standard
methods explained in Ref.~\cite{BILLO}, it is not difficult to check 
that the NS-NS contribution to $\Gamma$ is
\begin{eqnarray}
\Gamma_{\rm NS} &=& 
\frac{V_{p+1}}{2\pi}\,\frac{n^4}{\Delta_{m,n}}\,(8\pi^2\alpha')^{-\frac{p+1}{2}}
\int_{0}^\infty dt \left(\frac{\pi}{t}\right)^{\frac{9-p}{2}}
{\rm e}^{-r^2/(2\alpha't)} 
\nonumber \\
&&\times
\left(\frac{f_3^8(q)-f_4^8(q)}{f_1^8(q)}\right) 
\label{gammans}
\end{eqnarray}
where $q={\rm e}^{-t}$ and, as usual, 
\begin{eqnarray}
  \label{fi}
  f_1(q)=q^{{1\over 12}} \prod_{n=1}^\infty (1 - q^{2n}) ~~~~~~&,&~~
  f_2(q)=\sqrt{2}q^{{1\over 12}} \prod_{n=1}^\infty (1 + q^{2n}) ~~,
  \nonumber\\
  f_3(q)=q^{-{1\over 24}} \prod_{n=1}^\infty (1 + q^{2n -1})  ~~&,&~~
  f_4(q)=q^{-{1\over 24}} \prod_{n=1}^\infty (1 - q^{2n -1}) ~~.
\end{eqnarray}
Similarly, one can show that the R-R contribution to $\Gamma$
is 
\begin{equation}
\Gamma_{\rm R} = 
-\frac{V_{p+1}}{2\pi}\,\frac{n^4}{\Delta_{m,n}}\,(8\pi^2\alpha')^{-\frac{p+1}{2}}
\int_{0}^\infty dt \left(\frac{\pi}{t}\right)^{\frac{9-p}{2}}
{\rm e}^{-r^2/(2\alpha't)} 
~\frac{f_2^8(q)}{f_1^8(q)}~~.
\label{gammar}
\end{equation}
The ``abstruse identity'' satisfied by the $f$-functions implies that
$\Gamma=\Gamma_{\rm NS}+\Gamma_{\rm R} =0$, {\it i.e.} the BPS condition
at the full string level. 

We can therefore conclude that the boundary state $\ket{B}$ defined in
\secn{boundary} with an external electric field as in \eq{effe3}, really provides 
the complete conformal description of the BPS bound states formed
by fundamental strings and D$p$ branes.

\vskip 1.3cm
\sect{Dyonic strings in the Type IIB theory}
\label{Schwarz}
\vskip 0.5cm
In this section we consider in more detail the (F,D1) bound states
which describe the dyonic strings
first introduced by J. Schwarz in Ref.~\cite{SCHWARZ}. 
Let us recall that 
if we set to zero the self-dual five-form of the R-R sector, the low 
energy effective
action for the IIB string in the Einstein frame can be written as follows
\begin{equation}
S_{\rm IIB} = \frac{1}{ 2 \kappa^2} \int d^{10} x \sqrt{-g} \left[ R + 
\frac{1}{4}\,{\rm Tr}\left(\partial {\cal{M}} \,\partial {\cal{M}}^{-1}
\right) - 
\frac{1}{12} \,{\bf F}^{T} {\cal{M}} {\bf F}\right]
\label{IIB}
\end{equation}
where $R$ is the scalar curvature, ${\cal M}$ is the $SL(2,R)$ matrix
constructed out of the dilaton $\varphi$ and the axion $\chi$ according to
\begin{equation}
{\cal{M}} = {\rm e}^{\varphi} \left( \begin{array}{cc} |\lambda|^2 & \chi \\
                                           \chi & 1 
                          \end{array} \right)
\label{sl2rma}
\end{equation}
with
\begin{equation}
\lambda = \chi + \ii \,{\rm e}^{-\varphi}~~,
\label{lam}
\end{equation}
and finally ${\bf F}$ is the two-component vector ${\bf F}=(F_{\rm NS},F_{\rm R})$
formed by the field strengths $F_{\rm NS}=d{\widehat A}$
and $F_{\rm R}=d{\widehat C_{(2)}}$ of the two-form potentials of the
NS-NS and R-R sectors.
The action (\ref{IIB}) has manifest invariance under 
global $SL(2,R)$ transformations given by
\begin{equation}
{\cal{M}} ~\rightarrow ~\Lambda \,{\cal{M}}\, \Lambda^{T}
~~~~,~~~~
{\bf F} ~\rightarrow ~(\Lambda^T )^{-1}\, {\bf F}
\label{sl2rtra}
\end{equation}
where
\begin{equation}
\Lambda = \left( 
             \begin{array}{cc} a & b \\
                             c & d \end{array} \right) 
~~~~\mbox{with}~~ad-bc=1~~.
\label{lambda1}
\end{equation}
More explicitly, we have
\begin{equation}
\lambda ~\rightarrow ~\frac{a \lambda +b}{c \lambda +d} 
~~~,~~~
F_{\rm NS}~\rightarrow~ d\,F_{\rm NS} - c\,F_{\rm R} ~~~,~~~
F_{\rm R}~\rightarrow~ -b\,F_{\rm NS} + a\,F_{\rm R}~~.
\label{tras}
\end{equation}
The dyonic string is a classical solution of the field equations derived
from the action (\ref{IIB}) which is (electrically) charged under the two 
antisymmetric tensors of the NS-NS and R-R sectors. Let us observe
that if we denote by ${\bf J}_{\mu\nu}=({J_{\rm NS}}_{\mu\nu},
{J_{\rm R}}_{\mu\nu})$ the current which is coupled to the antisymmetric
tensors, then the charge ${\bf q}=(q_{\rm NS},q_{\rm R})$
of the dyonic string is defined by 
\begin{eqnarray}
{\bf q} &\equiv & \int_{V_8}d^8x_\bot \,{\bf J}_{01} =
\frac{1}{\sqrt{2}\kappa}\,\int_{V_8}
d^8x_\bot\,\partial_\mu
\left [ \sqrt{-g}\,{\cal M}\,{\bf F}_{01}^{~~\mu}\right] 
\label{qua4.51} \\
&=& \frac{1}{\sqrt{2}\kappa}\,\frac{1}{7!}\,
\varepsilon_{\mu_1 \dots \mu_7 \mu}
\int_{\partial V_8}\,dx^{\mu_1} \wedge 
\dots \wedge dx^{\mu_7} \,
\left [ \sqrt{-g}\,{\cal M}\,{\bf F}_{01}^{~~\mu}\right]_{\infty}
\nonumber
\end{eqnarray}
where in the last step we have used Stoke's theorem. In these formulas
$V_8$ denotes the space transverse to the string world-sheet whose 
boundary $\partial V_8$ is a seven dimensional sphere at infinity.
{F}rom Eqs. (\ref{qua4.51}) and (\ref{sl2rtra}) it is easy to realize 
that the electric charges transform under an $SL(2,R)$ transformation 
$\Lambda$ according to
\begin{equation}
\label{chargetra}
{\bf q} ~ \rightarrow ~ \Lambda\,{\bf q}~~.
\end{equation}
As is well known, not all classical solutions are fully consistent at the
quantum level; only 
those which carry {\it integer} charges in units of $\mu_1$ (see
\eq{cou89} for $p=1$) satisfy the Dirac quantization condition and are
acceptable. 
This is precisely the case of the (F,D1) bound state
discussed in \secn{fdp}. For later convenience we write explicitly
the long distance behavior of the corresponding fields obtained
from the boundary state and given in Eqs. (\ref{dilft})-(\ref{cp1ft}) for 
$p=1$, namely
\begin{equation}
\label{scalsch}
\delta\varphi = - \frac{m^2-n^2}{2\,\Delta_{m,n}}\,\frac{Q_1}{r^6}
~~~,~~~\delta\chi =  \frac{mn}{\Delta_{m,n}}\,\frac{Q_1}{r^6}~~,
\end{equation}
\begin{equation}
\label{metsch}
\delta g_{\mu\nu} = {\rm 
diag}\left(\frac{3}{4},-\frac{3}{4},\frac{1}{4},\cdots,\frac{1}{4}
\right)\,\frac{Q_1}{r^6}~~,
\end{equation} 
and
\begin{equation}
\label{antissch}
\delta{\widehat A}_{01}= -\frac{m}{\Delta_{m,n}^{1/2}}\,\frac{Q_1}{r^6}
~~~,~~~\delta{\widehat C}_{01} = 
-\frac{n}{\Delta_{m,n}^{1/2}}\,\frac{Q_1}{r^6}~~.
\end{equation}
where $Q_1$ is given in \eq{Qp} for $p=1$.
Using \eq{antissch} into \eq{qua4.51}, and remembering that in this case
the asymptotic matrix ${\cal M}|_{\infty}$ is the identity, one
can easily verify that
\begin{equation}
\label{charges}
q_{\rm NS} = m\,\mu_1~~~,~~~q_{\rm R}= n\,\mu_1~~,
\end{equation}
that is, as expected, the two charges 
are integer multiples of $\mu_1$.

We now want to generalize these considerations to the case
in which the two scalar fields $\varphi$ and $\chi$ have non
vanishing asymptotic values $\varphi_0$ and $\chi_0$ respectively.
This can be easily achieved by exploiting the $SL(2,R)$
invariance of the theory and ``rotating'' the solution
given in Eqs. (\ref{scalsch})-(\ref{antissch}) by means
of the following transformation 
\begin{equation}
\Lambda = \left( 
             \begin{array}{cc} e^{-\varphi_0 /2} &  \chi_0 \,e^{\varphi_0 
/2} \\ 0 & e^{\varphi_0 /2} \end{array} \right)~~.
\label{trasfs}
\end{equation}
Indeed, according to \eq{tras} we have
\begin{equation}
\label{newfields}
\varphi ~ \rightarrow ~ {\widetilde\varphi} = \varphi +\varphi_0
~~~~,~~~~
\chi ~ \rightarrow ~ {\widetilde \chi} = {\rm e}^{-\varphi_0}\,\chi + \chi_0
~~,
\end{equation}
so that the transformed fields acquire the desired asymptotic values. Furthermore,
under $\Lambda$ the antisymmetric tensors transform as follows 
\begin{equation}
\label{newfields1}
{\widehat A}_{01} ~\rightarrow~ {\widetilde A}_{01}=
{\rm e}^{\varphi_0/2}\,{\widehat A}_{01}
~~~~,~~~~
{\widehat C}_{01} ~\rightarrow~ {\widetilde C}_{01}=
{\rm e}^{-\varphi_0/2}\,{\widehat C}_{01} - \chi_0\,{\rm e}^{\varphi_0/2}
\,{\widehat A}_{01}~~,
\end{equation}
and correspondingly the charges become
\begin{equation}
\label{chargetra1}
q_{\rm NS} ~\rightarrow~{\widetilde q}_{\rm NS} = 
{\rm e}^{-\varphi_0/2}\,q_{\rm NS} + 
\chi_0\,{\rm e}^{\varphi_0/2}\,q_{\rm R}
~~~~,~~~~
q_{\rm R}~\rightarrow~{\widetilde q}_{\rm R} = {\rm e}^{\varphi_0/2}
\,q_{\rm R}~~. 
\end{equation}
The new configuration is acceptable only if the new
charges ${\widetilde q}_{\rm NS}$ and ${\widetilde q}_{\rm R}$
obey the Dirac quantization condition. 
{F}rom \eq{chargetra1} we easily see that this
condition is realized if we start from a configuration like
the one of \eq{antissch} but with $m$ and $n$ replaced according to
\begin{equation}
m ~\rightarrow ~ {\rm e}^{\varphi_0 /2} \,(m - \chi_0 \,n )~~~,~~~
n ~\rightarrow ~{\rm e}^{- \varphi_0 /2}\,n~~.
\label{q2}
\end{equation}
In view of the discussion of \secn{fdp}, we can say that this
configuration can be obtained from a 
boundary state with an external field on it given by
\begin{equation}
{\widetilde{f}} = -
\frac{{\rm e}^{\varphi_0 /2} \, ( m - 
\chi_0\,n)}{{\widetilde \Delta}_{m,n}^{1/2}} 
\label{efe4}
\end{equation}
where
\begin{equation}
\label{delta1}
{\widetilde \Delta}_{m,n} = {\rm e}^{\varphi_0} \,(m - \chi_0 \,n )^2
+{\rm e}^{- \varphi_0}\,n^2~~.
\end{equation}
Notice that these expressions are simply obtained from Eqs. (\ref{xf3})
and (\ref{Delta}) with the substitutions (\ref{q2}). Furthermore, we remark
that ${\widetilde f}$ is precisely the gauge field found in 
Ref.~\cite{SCHMID} with different considerations. 

Performing the transformations on the massless fields
as indicated in Eqs. (\ref{newfields}) and (\ref{newfields1}),
we finally obtain the following long distance behavior for the scalars
\begin{equation}
\label{newscal}
\delta{\widetilde\varphi}
= - \frac{{\rm e}^{\varphi_0}\,(m-\chi_0\,n)^2
-{\rm e}^{-\varphi_0}n^2}{2\,{\widetilde\Delta}_{m,n}}
\,\frac{{\widetilde Q}_1}{r^6} ~~~,~~~
\delta{\widetilde\chi} = 
\frac{{\rm e}^{-\varphi_0}\,
(m-\chi_0\,n)\,n}{{\widetilde\Delta}_{m,n}}\,
\frac{{\widetilde Q}_1}{r^6 }~~,
\end{equation} 
and for the antisymmetric tensors
\begin{equation}
\label{newantis}
\delta{\widetilde A}_{01}=
-\frac{{\rm 
e}^{\varphi_0}\,(m-\chi_0\,n)}{{\widetilde\Delta}_{m,n}^{1/2}}
\,\frac{{\widetilde Q}_1 }{r^6} 
~~~,~~~
\delta{\widetilde C}_{01} = 
-\frac{{\rm e}^{\varphi_0}\,(|\lambda_0|^2\,n
-\chi_0\,m)}{{\widetilde\Delta}_{m,n}^{1/2}}
\,\frac{{\widetilde Q}_1}{r^6}
\end{equation}
where $\widetilde Q_1$ is defined as in \eq{Qp} with
$\widetilde \Delta_{m,n}$ in place of $\Delta_{m,n}$.
These expressions are in complete agreement with the long distance
behavior of the dyonic string solution
of Schwarz \cite{SCHWARZ}. Thus, we can conclude that a boundary state with
an external electric field $\widetilde f$ represents the exact conformal description
of the dyonic configurations with arbitrary background values
at the full string level.

\vskip 1.3cm
\sect{Concluding remarks}
\label{conclusions}
\vskip 0.5cm

In this paper we have shown that the boundary state with a constant
electric field provides the complete conformal representation for the BPS
bound states formed by a fundamental string and a D$p$ brane.
At the classical level, these configurations interpolate between the 
pure fundamental string and the pure D$p$ brane. In fact the latter can
be obtained by setting $f=0$ (or equivalently $m=0$),
while the fundamental string is realized by choosing $f=-1$ (or 
equivalently $n=0$) in Eqs.(\ref{dilft})-(\ref{cp1ft}).

One may wonder whether the same interpolation can be done at the full 
conformal level, {\it i.e.}
directly on the boundary state. While there is of 
course no problem in switching off the electric field to obtain the 
boundary state for a pure D$p$ brane, the other limit, $f \to -1$ (or 
equivalently $n \to 0$), is singular and not well defined on the 
boundary state.
The reason is that the Dirac-Born-Infeld prefactor $n \sqrt{1-f^2}$
vanishes in this limit, while the longitudinal part of the $S$ matrix 
(see Eq. (\ref{emme3})) diverges.
However, these two effects compensate each other whenever the boundary 
state is saturated with closed string states which contain at most one 
left and one right oscillator with longitudinal index. This is precisely
the structure of the massless states of the NS-NS sector.
Using this observation and the fact that the R-R sector does not play 
any role in the fundamental string solution, we can introduce an
effective operator which generates the classical fundamental string in
the same way as the boundary state does for the D brane. 
Such operator is equal in structure to a boundary state without R-R 
sector and with a NS-NS sector defined by a longitudinal $(2 \times 
2)$ $S$ matrix given by \begin{equation}
S_{\alpha\beta} =  
\left(\begin{array}{cc} 
           -2  & -2 \\
            2  & 2 
          \end{array} \right)
\label{Sf}
\end{equation}
and no transverse $S$ matrix.
This effective operator is perfectly well defined and, when it is
projected onto the dilaton, the graviton and the
antisymmetric tensor, it yields the fundamental string solution.
Furthermore, it appears to describe 
a BPS object, because the vacuum amplitude between two such 
states is identically vanishing. Therefore, it can be considered as 
an effective conformal representation of the fundamental
string which is valid at very large distances, {\it
i.e.} only for the massless fields. However, it cannot be the complete
conformal description of the fundamental string solution.
In fact, a ``coherent state " structure, like the one given for instance
in Eq. (\ref{bs100}), always enforces an identification between
left and right oscillators, which is appropriate only when the string 
world sheet has a boundary; this is certainly not the case for the pure 
fundamental string which only couples to closed string states and 
therefore must have independent left and right sectors.
Finding the complete conformal description of the fundamental string
remains therefore an open problem.

\vskip 1.3cm
{\large{\bf Acknowledgements }}
\vskip .5cm
\noindent
One of the authors (A. Lic.) thanks R. Marotta for very useful discussions and 
NORDITA for the kind hospitality. We also thank R. Russo for many
interesting discussions. 

\vskip 1.3cm
\appendix{\Large {\bf{Appendix A}}}
\label{appe}
\vskip 0.5cm
\renewcommand{\theequation}{A.\arabic{equation}}
\setcounter{equation}{0}
\noindent
In this appendix we give some details about the 
massless closed string states and define the corresponding projectors.
In the $(-1,-1)$ superghost picture of the NS-NS sector, 
the massless states are 
\begin{eqnarray}
\ket{\phi} &=&  \frac{\phi}{\sqrt{8}} \left( \eta_{\mu \nu} - 
k_{\mu} \ell_{\nu} -  k_{\nu} \ell_{\mu}\right)\,
{\tilde{\psi}}^{\mu}_{-\frac{1}{2}} \psi^{\nu}_{-\frac{1}{2}} 
~\ket{k/2}_{-1}\, \widetilde{\ket{k/2}}_{-1}~~,
\label{dila} \\
\ket{h} &=& h_{\mu \nu}~{\tilde{\psi}}^{ \mu}_{-\frac{1}{2}} 
\psi^{\nu}_{-\frac{1}{2}}~
\ket{k/2}_{-1}\, \widetilde{\ket{k/2}}_{-1}~~,
\label{grav}\\
\ket{A} &=&\frac{A_{\mu \nu}}{\sqrt{2}}~{\tilde{\psi}}^{ \mu}_{-\frac{1}{2}} 
\psi^{ \nu}_{-\frac{1}{2}}~\ket{k/2}_{-1}\, \widetilde{\ket{k/2}}_{-1}
\label{anti}
\end{eqnarray} 
where $\phi$, $h_{\mu\nu}$ and $A_{\mu\nu}$
are the dilaton, graviton and Kalb-Ramond polarizations
respectively.
The corresponding conjugate states are 
\begin{eqnarray}
\bra{\phi}  &=&  {}_{-1} \widetilde{\bra{k/2}}~ 
{}_{-1}\bra{k/2} ~  
\psi^{ \nu}_{\frac{1}{2}} {\tilde{\psi}}^{ \mu}_{\frac{1}{2}} 
~\frac{\phi}{\sqrt{8}} \left( \eta_{\mu \nu} - 
k_{\mu} \ell_{\nu} - k_{\nu} \ell_{\mu} \right)~~,
\label{dila1}\\
\bra{h} &=& {}_{-1} \widetilde{\bra{k/2}} ~
{}_{-1}\bra{k/2}~\psi^{\nu}_{\frac{1}{2}}
 {\tilde{\psi}}^{ \mu}_{\frac{1}{2}} ~h_{\mu\nu}~~,
\label{grav1} \\
\bra{A} &=& {}_{-1} \widetilde{\bra{k/2}}~{}_{-1}\bra{k/2}  
~\psi^{ \nu}_{\frac{1}{2}} {\tilde{\psi}}^{\mu}_{\frac{1}{2}} 
~\frac{A_{\mu \nu}}{\sqrt{2}}~~. 
\label{anti1}
\end{eqnarray}
The normalization of these states 
has been chosen in such a way that their norms 
\beq
\braket{\phi}{\phi} = \phi^2
~~~,~~~
\braket{h}{h} =  h^{\mu \nu}\,h_{\mu \nu}
~~~,~~~
\braket{A}{A} = \frac{1}{2} \,A^{\mu \nu} A_{\mu \nu} 
\label{normali}
\eeq
correspond to canonically normalized fields.

It is useful to define also projection operators that, when
applied to an arbitrary massless state of the closed string,
select the graviton, the dilaton and the Kalb-Ramond field
components contained in that state.
Given the form of the
massless states (\ref{dila})-(\ref{anti1}), it is not
difficult to verify that these projectors 
are 
\bea
\bra{P^{(\phi)}} &=&  {}_{-1} 
\widetilde{\bra{k/2}}~{}_{-1} \bra{k/2}
~\psi^{\nu}_{\frac{1}{2}} {\tilde{\psi}}^{\mu}_{\frac{1}{2}}~
\frac{1}{\sqrt{8}}\left( \eta_{\mu \nu} -  k_{\mu} \ell_{\nu} - 
k_{\nu} \ell_{\mu} \right)~~,
\label{dila45}\\
\bra{{P^{(h)}}^{\mu \nu}} &=& 
{}_{-1} \widetilde{\bra{k/2}}~
{}_{-1}\bra{k/2}~\frac{1}{2}\left( \psi^{\nu}_{\frac{1}{2}}
{\tilde{\psi}}^{\mu}_{\frac{1}{2}} +
\psi^{\mu}_{\frac{1}{2}} 
{\tilde{\psi}}^{\nu}_{\frac{1}{2}}\right) 
\nonumber \\
&&~~-~ \bra{P^{(\phi)} }~ 
\frac{1}{\sqrt{8}}
\left(\eta^{\mu \nu} -  k^{\mu} \ell^{\nu} - 
k^{\nu} \ell^{\mu} \right)~~,
\label{gravi87} \\
\bra{ {P^{(A)}}^{\mu \nu}} &=& {}_{-1} 
\widetilde{\bra{k/2}}~{}_{-1}\bra{k/2}
~\frac{1}{\sqrt{2}} \left( \psi^{\nu}_{\frac{1}{2}} 
{\tilde{\psi}}^{\mu}_{\frac{1}{2}} -
\psi^{\mu}_{\frac{1}{2}} 
{\tilde{\psi}}^{\nu}_{\frac{1}{2}} \right)~~.
\label{anti76} 
\ena
Indeed, they satisfy the following properties
\beq
\braket{P^{(\phi)}}{\phi} = \phi 
~~~,~~~
\braket{P^{(h)}_{\mu\nu}}{h} = h_{\mu\nu}
~~~,~~~
\braket{P^{(A)}_{\mu\nu}}{A} = A_{\mu \nu}~~~,
\label{pola54}
\eeq
with all other overlaps being zero.

Let us now consider the massless states of the R-R sector.
As we mentioned in \secn{boundary}, in order to have a non vanishing
overlap with the boundary state, we must work in the asymmetric
$(-1/2,-3/2)$ picture so that the superghost number anomaly
of the disk is soaked up. In \eq{fred6t} we wrote an
expression for the state representing a $n$-form R-R
potential $C_{(n)}$. However, that expression has to be interpreted
as an ``effective'' and simplified description which can be used 
only when parity violating terms do not contribute to scattering
amplitudes.
The complete expression for the massless R-R states instead involves
infinite terms with different superghost number which combine to
give \cite{BILLO}
\begin{eqnarray}
\label{fred6}
\ket{C_{(n)}} & = & \frac{1}{2\sqrt{2}\,n!}~
C_{\mu_1\ldots\mu_n}\Bigg[
\left(C\Gamma^{\mu_1\ldots\mu_n}\Pi_+\right)_{AB}\,
\cos (\gamma_0{\widetilde\beta}_0) 
\\
&&~+~
\left( C\Gamma^{\mu_1\ldots\mu_n}\Pi_-\right)_{AB}\,
\sin (\gamma_0{\widetilde\beta}_0) \Bigg]
\ket{A;k/2}_{-1/2}~\ket{{\widetilde B};\widetilde{k/2}}_{-3/2}
\nonumber
\end{eqnarray}
where $\Pi_\pm = (1\pm\Gamma_{11})/2$, and $\beta_0$ and $\gamma_0$
are the superghost zero-modes. The corresponding conjugate state is
\begin{eqnarray}
\bra{C_{(n)}} &=&
{}_{-1/2}\bra{\widetilde{B}, \widetilde{k/2}} ~{}_{-3/2}\bra{A, k/2} 
\Bigg[\left(C\Gamma^{\mu_1\ldots\mu_n}\Pi_-\right)_{AB}
\cos ( \beta_0 {\widetilde{\gamma}}_0) 
\nonumber \\
&&+
\left(C\Gamma^{\mu_1\ldots\mu_n}\Pi_+\right)_{AB}
\sin( \beta_0 {\widetilde{\gamma}}_0)
\Bigg]\,\frac{(-1)^n}{2\sqrt{2}\,n!} \,C_{\mu_1\ldots\mu_n}~~.
\label{sta}
\end{eqnarray}
We now show that these states are correctly normalized. Due to the 
presence of the superghost zero modes, the norm
$\braket{C_{(n)}}{C_{(n)}}$ is naively divergent and a suitable regularization
is necessary \cite{YOST,BILLO}. This can be performed by inserting
in the scalar product the operator
$x^{2F_0-2\gamma_0\beta_0}$, where $F_0$ is the zero-mode part of the
world-sheet fermion number, and letting $x\to 1$ at the end.
Keeping this in mind, let us first compute the superghost 
contribution. Using the equation
\beq
\bra{\widetilde{-1/2}} ~\bra{{-{3/2}}}
\,{\rm  e}^{\ii\eta_1\beta_0\tilde\gamma_0}~x^{-2\gamma_0\beta_0}~
  {\rm e}^{\ii\eta_2\gamma_0\tilde\beta_0}\ket{-{1/2}}~ 
  \ket{\widetilde{-{3/2}}}=
\frac{1}{1 - \eta_1 \eta_2 x^2}~~,
\label{bs50}
\eeq
we can easily obtain
\bea
J &\equiv& \bra{\widetilde{-{1/2}}}~\bra{-{3/2}}
\cos (\beta_0\tilde\gamma_0)~x^{-2\gamma_0\beta_0}\,
\cos (\gamma_0{\tilde{\beta}}_{0})\, \ket{-{1/2}} ~\ket{\widetilde{-{3/2}}}
\nonumber \\
&=& \frac{1}{2} \left( \frac{1}{1 - x^2}+ \frac{1}{1 + x^2} \right)~~,
\label{bs51}
\\
K &\equiv& \bra{\widetilde{-{1/2}}} ~\bra{-{3/2}} 
\sin (\beta_0 {\tilde{\gamma}}_{0}) ~x^{-2\gamma_0\beta_0}\,
\sin (\gamma_0\tilde\beta_0)\,\ket{-{1/2}} ~\ket{\widetilde{-{3/2}}} =
\nonumber \\
&=&- \frac{1}{2} \left( \frac{1}{1 -  x^2}-  \frac{1}{1 +  x^2} 
\right)~~,
\label{bs78}
\ena
and also check that analogous expressions with one sine and one
cosine are vanishing. Then, recalling that the fermionic vacua are such that
$\braket{A}{B}=\braket{\widetilde A}{\widetilde B}=(C^{-1})^{AB}$, we get
\bea
\braket{C_{(n)}}{C_{(n)}}
&=& \frac{(-1)^n}{8\,(n!)^2}~C_{\mu_1\ldots\mu_{n}}\, 
C_{\nu_1\ldots\nu_{n}}
\nonumber \\
&\times &\lim_{x\to 1}\left\{ J\,{\rm Tr} \left[ x^{2F_0}
\left( C \Gamma^{\mu_1\ldots\mu_{n}} \Pi_- C^{-1} 
\right)^{\rm T}\Gamma^{\nu_1\ldots\nu_{n}} \Pi_+ 
\right]\right.
\nonumber \\
&+& \left. K \,{\rm Tr}  
\left[x^{2F_0}\left( C \Gamma^{\mu_1\ldots\mu_{n}} 
\Pi_+ C^{-1} \right)^{\rm T}\Gamma^{\nu_1\ldots\nu_{n}} 
\Pi_- \right]\right\}~~.
\label{equa32} 
\ena
Using the transposition properties of the $\Gamma$ matrices
and exploiting Eqs. (\ref{bs51}) and (\ref{bs78}), after some
simple algebra we obtain
\begin{eqnarray}
\braket{C_{(n)}}{C_{(n)}}  &=& \frac{1}{16\,(n!)^2}~C_{\mu_1\ldots 
\mu_{n}} \,C_{\nu_1\ldots \nu_{n}} \lim_{x\to 1}
\left[{\rm Tr}\left(x^{2F_0}  \Gamma^{\mu_{n} 
\ldots\mu_1} \Gamma^{\nu_1\ldots\nu_{n}} \right)
\frac{1}{1+x^2}\right.
\nonumber \\
&+& \left. {\rm Tr}\left(
x^{2F_0} \Gamma^{\mu_{n}
\ldots\mu_1} \Gamma^{\nu_1\ldots\nu_{n}} \Gamma_{11}\right)
\frac{1}{1-x^2}\right]~~.
\label{equa54}
\ena
The parity violating term containing $\Gamma_{11}$ is
identically zero, so that we can safely take the limit $x\to 1$
and get
\begin{eqnarray}
\braket{C_{(n)}}{C_{(n)}} &=&  \frac{1}{32\,(n!)^2}
 ~C_{\mu_1 \ldots \mu_{n}}
\,C_{\nu_1 \ldots \nu_{n}}\,{\rm  Tr} \left( \Gamma^{\mu_{n}
\ldots\mu_1} \Gamma^{\nu_1\ldots\nu_{n}} \right)
\nonumber \\
&=& \frac{1}{n!}~
C^{\mu_1 \ldots \mu_{n}} C_{\mu_1 \ldots \mu_{n}} 
\label{equa53}~~.
\end{eqnarray}
This result shows that the state (\ref{fred6}) is canonically normalized.
Now we can define the projection operator associated to $\ket{C_{(n)}}$
which turns out to be
\begin{eqnarray}
\bra{P^{(C)}_{\mu_{1} \ldots \mu_{n}}} &=&
{}_{-1/2}\bra{\widetilde{B}, \widetilde{k/2}} ~{}_{-3/2}\bra{A, k/2} 
\Bigg[\left(C\Gamma_{\mu_1\ldots\mu_n}\Pi_-\right)_{AB}
\cos ( \beta_0 {\widetilde{\gamma}}_0) 
\nonumber \\
&&+
\left(C\Gamma_{\mu_1\ldots\mu_n}\Pi_+\right)_{AB}
\sin( \beta_0 {\widetilde{\gamma}}_0) 
\Bigg]\,\frac{(-1)^n}{2\sqrt{2}}~~. \label{sta65}
\end{eqnarray}
Repeating the same steps described before, one can check that indeed
\begin{equation}
\braket{P^{(C)}_{\mu_{1} \ldots \mu_{n}}}{C_{(n)}} = C_{\mu_1 \ldots \mu_{n}}~~.
\label{proi87}
\eeq
Let us now compute the overlap of $\bra{P^{(C)}}$ with the R-R boundary state.
Since the projector (\ref{proi87}) contains only zero-modes, it is 
enough to consider the zero-mode part of the R-R boundary state, which
is explicitly given by \cite{BILLO}
\begin{equation}
\label{brr00}
\ket{B,\eta}_{\rm R}^{(0)} =-\frac{T_p}{2}~\delta^{(9-p)}(q^i-y^i)~
{\rm e}^{\ii\,\eta\,\gamma_0\,{\widetilde\beta}_0}~
{\cal M}^{(\eta)}_{AB}~
\,\ket{A,0}_{-1/2} \,\ket{{\widetilde B},\widetilde 0}_{-3/2}
\end{equation}
where 
\beq
{\cal{M}}^{(\eta)}_{AB} = \left[C \Gamma^{0} \cdots \Gamma^{p} 
\frac{1+\ii \eta \Gamma_{11}}{1+ \ii \eta} \,
;{\rm e}^{ - 1/2 {\hat{F}}_{\alpha \beta} \Gamma^{\alpha}
\Gamma^{\beta}}  ; \right]_{AB}
\label{mab4}
\eeq
Then, using \eq{sta65}, we obtain
\begin{eqnarray}
\braket{P^{(C)}_{\mu_1\ldots\mu_n}}{B,\eta}_{\rm R}
& = & 
-V_{p+1}\,T_p\,\frac{(-1)^{n}}{4 \sqrt{2}} ~\lim_{x\to 1}~\Bigg\{
{}_{-1/2}\bra{\widetilde{D}} ~{}_{-3/2}\bra{C} 
~\times
\nonumber \\
&&\left[\left(C\Gamma_{\mu_1\ldots\mu_n}\Pi_-\right)_{CD}
\cos ( \beta_0 {\widetilde{\gamma}}_0) +
\left(C\Gamma_{\mu_1\ldots\mu_n}\Pi_+\right)_{CD}
\sin ( \beta_0 {\widetilde{\gamma}}_0) 
\right]
\nonumber \\
&&x^{2F_0-2\gamma_0\beta_0}~
{\cal M}^{(\eta)}_{AB}~{\rm e}^{\ii\,\eta\gamma_0{\widetilde \beta}_0}
~\ket{A}_{-1/2} \ket{\widetilde{B}}_{-3/2}\Bigg\}~~.
\label{mat78}
\end{eqnarray}
Let us compute first the superghost contribution. 
Using \eq{bs50}, we get
\begin{eqnarray}
\bra{\widetilde{-{1/2}}}\,\bra{-{3/2}}
\cos ( \beta_0\tilde\gamma_0 )\,x^{-2\gamma_0\beta_0}\,
{\rm e}^{\ii\,\eta\gamma_0{\widetilde\beta}_0}
\,\ket{-{1/2}}\,\ket{\widetilde{-{3/2}}}
\!&=&\! \frac{1}{2}
\left[ \frac{1}{1 - \eta x^2} + \frac{1}{1 + \eta x^2}\right]~,
\nonumber \\
\bra{\widetilde{-{1/2}}}\,\bra{-{3/2}}
\sin ( \beta_0\tilde\gamma_0 )\,x^{-2\gamma_0\beta_0}\,
{\rm e}^{\ii\,\eta\gamma_0{\widetilde\beta}_0}
\,\ket{-{1/2}}\,\ket{\widetilde{-{3/2}}}
\!&=& \!\frac{1}{2\ii}
\left[ \frac{1}{1 - \eta x^2} - \frac{1}{1 + \eta x^2}\right]~;
\nonumber 
\end{eqnarray}
then, after some simple manipulations, \eq{mat78} becomes
\begin{eqnarray}
&&-V_{p+1}\,T_p\,\frac{1}{8\sqrt{2}}   
\left\{ {\rm Tr} \left(x^{2F_0}\,\Pi_+ \Gamma_{\mu_{n}\ldots\mu_1} C^{-1} 
{\cal{M}}^{(\eta)}  \right) \left( \frac{1}{1- \eta x^2} + 
\frac{1}{1+ \eta x^2} \right) +\right.
\nonumber \\
&&~~~~\left. -\ii\, {\rm Tr} \left(x^{2F_0}\,
\Pi_- \Gamma_{\mu_{n} \ldots \mu_1} C^{-1} 
{\cal{M}}^{(\eta)}  \right)  \left( \frac{1}{1- \eta x^2} - 
\frac{1}{1+ \eta x^2} \right) \right\}
~~.
\label{equ52}
\end{eqnarray}
Finally, summing over the 
two R-R spin-structures to perform the GSO projection, we get
\begin{eqnarray}
\braket{P^{(C)}_{\mu_1\ldots\mu_n}}{B}_{\rm R}
&=& -V_{p+1}\,T_p\,\frac{1-(-1)^{n+p}}{16\sqrt{2}} \\
&\times &
\lim_{x\to 1}\Bigg\{
{\rm Tr} \left(x^{2F_0}\,\Gamma_{\mu_n \ldots \mu_1} 
\Gamma^0\cdots \Gamma^p 
~; {\rm e}^{-1/2 {\hat{F}}^{\alpha \beta} \Gamma_{\alpha} \Gamma_{\beta}} ;   
\right)
\frac{1}{1+x^2} 
\nonumber \\
&&+~
{\rm Tr} \left(x^{2F_0}\Gamma_{\mu_{n}\ldots\mu_1} 
\Gamma^0\cdots\Gamma^p\Gamma_{11}~
; {\rm e}^{-1/2 {\hat{F}}^{\alpha \beta} \Gamma_{\alpha} \Gamma_{\beta}} ;
\right)
\frac{1}{1-x^2} 
\Bigg\}~~.
\nonumber 
\label{equ53}
\end{eqnarray}
If the indices $\mu_1\ldots\mu_n$ are all along the world-volume
of the D$p$-brane (which is precisely the case of interest in
this paper), then the term containing $\Gamma_{11}$ in Eq.(\ref{equ53}) 
vanishes; therefore, taking the limit $x\to 1$ and observing that 
$n$ and $p$ must have opposite parity in order to have a non-zero
result, we finally obtain
\begin{equation}
\braket{P^{(C)}_{\mu_1\ldots\mu_n}}{B}_{\rm R} = 
-V_{p+1} \,T_p\,\frac{1}{16 \sqrt{2} } 
{\rm Tr} \left( \Gamma_{\mu_n \ldots \mu_1} \Gamma^0 \dots \Gamma^p~
 ;{\rm e}^{ - 1/2  {\hat{F}}_{\alpha \beta} \Gamma^{\alpha} \Gamma^{\beta}}  
;\right)~~.
\label{wnplus}
\end{equation}
Taking into account the relation 
\begin{equation}
\label{reprosta}
\bra{P^{(C)}_{\mu_1\ldots\mu_n}}\, 
\frac{C^{\mu_1 \dots \mu_n}}{n!} = \bra{C^{(n)}}~~,
\end{equation}
we see that \eq{wnplus} reproduces \eq{wnplus1}.

In the final part of this appendix we explicitly write the form 
of the conjugate
boundary state that has to be used in \eq{vacampl}. For the bosonic 
coordinate we have
\begin{equation} 
\label{bs1001} 
\bra{B_X} = ~{}^{(0)}\bra{B_X}\,
\exp\biggl[-\sum_{n=1}^\infty \frac{1}{n}\, \a_{n}\cdot S\cdot
\tilde \a_{n} \biggr]~~,
\end{equation}
while for the fermionic part we have
\begin{equation}
\label{bs1011}
{}_{\rm NS}\bra{B_\psi,\eta} =  \ii\,\bra{0}\, 
\exp\biggl[- \ii\,\eta\sum_{m=1/2}^\infty
\psi_{m}\cdot S\cdot\tilde \psi_{m}\biggr] 
\end{equation}
in the NS-NS sector, and
\begin{equation}
\label{bs1021}
{}_{\rm R}\bra{B_\psi,\eta}= - ~{}_{\,\rm R}^{(0)}\!\bra{B,\eta} 
\,\exp\biggl[- \ii\,\eta\sum_{m=1}^\infty
\psi_{m} \cdot S\cdot\tilde \psi_{m}\biggr]
\end{equation}
in the R-R sector, where
\beq
{}_{\,\rm R}^{(0)}\!\bra{B,\eta} =  (-1)^p \bra{A} \,\bra{\widetilde B}
\left( C\Gamma^0\Gamma^{1}\ldots \Gamma^{p} \,
\frac{1-\ii\eta\Gamma_{11}}{1+\ii\eta}\, U\right)_{AB}~~
\label{brazemo}
\eeq
with $U$ given in \eq{umat}.
\vskip 1.3cm
\appendix{\Large {\bf{Appendix B}}}
\label{appb}
\vskip 0.5cm
\renewcommand{\theequation}{B.\arabic{equation}}
\setcounter{equation}{0}
\noindent
In this appendix we show how to derive from the boundary
state the classical solution
corresponding to the bound state
(W,D$p$) between a Kaluza-Klein wave W and a D$p$ brane.
The (W,D$p$) bound state can be easily obtained from a 
(F,D$_{p+1}$) configuration
by performing a T-duality along the
longitudinal spatial direction of the fundamental string ({\it i.e.} 
along $X^1$ in our conventions). As shown in Ref.~\cite{cpb},
a T-duality transformation in a given direction 
is simply realized by changing the sign in the corresponding row of
the $S$ matrix which defines the boundary state. Therefore, in our case, 
we can start from the $(p+2)\times (p+2)$ longitudinal $S$ matrix of the 
(F,D$(p+1)$) configuration given in \eq{emme3}, and then change the sign
of the second row corresponding to $X^1$ to get
\begin{equation}
S_{\alpha\beta} =  
\left( \begin{array}{ccccc} 
           -\frac{1 + f^2}{1-f^2}  & \frac{2f}{1-f^2}& & &  \\
                        \frac{2f}{1-f^2} & -\frac{1+f^2}{1-f^2} & & & \\
			 & & 1 & &  \\
			 & & &\ddots &  \\
			 & & & & 1
			 \end{array} \right)~~.
\label{emme3b}
\end{equation}
Notice that the matrix (\ref{emme3b}) is symmetric, as opposed to the 
one of \eq{emme3} which is antisymmetric. 

Following the same procedure described in \secn{fdp}, we 
can obtain the long distance behavior of the massless fields emitted by 
the (W,D$p$) bound state. For simplicity, we will concentrate
on the massless fields of the NS-NS sector. When written in terms of 
$T_{\mu\nu}$ (see \eq{flu34}), the long distance behavior
of these fields is formally equal to the one of the 
(F,D$_{p+1}$) bound state and is given by Eqs. (\ref{dil45}), 
(\ref{anti67}) and (\ref{gra3}).
However, since the new matrix $T_{\mu\nu}$ is symmetric, we can 
immediately conclude the no Kalb-Ramond field is emitted by the 
(W,D$p$) bound state. On the contrary, there is an off-diagonal 
component in $T_{\mu\nu}$ which gives rise to an off-diagonal component
in the metric. This is the distinctive feature of this configuration.

Writing the boundary state with the matrix (\ref{emme3b})
where the external field $f$ is given by \eq{xf3}, 
and performing a Fourier transformation, we then get
the following long distance behavior for the dilaton 
\begin{equation}
\varphi 
\simeq  \frac{3-p}{4}\,\frac{n^2}{\Delta_{m,n}}~\frac{Q_{p+1}}{r^{6-p}}~~,
\label{dilftb}
\end{equation}
and for the metric components
\begin{eqnarray}
g_{00} &\simeq & -1 -
\left(-1+\frac{p+1}{8}\,\frac{n^2}{\Delta_{m,n}}\right)\,
\frac{Q_{p+1}}{r^{6-p}}~~,
\nonumber \\
g_{01} &=&g_{10} \simeq  -\frac{m}{\Delta^{1/2}_{m,n}}
~\frac{Q_{p+1}}{r^{6-p}}~~,
\nonumber \\
g_{11} &\simeq & 1 + \left(1+\frac{p-7}{8}\,\frac{n^2}{\Delta_{m,n}}\right)\,
\frac{Q_{p+1}}{r^{6-p}}~~,
\nonumber \\
g_{22}&=&\ldots~=~g_{pp} \simeq
1 + \frac{p-7}{8}\,\frac{n^2}{\Delta_{m,n}}~
\frac{Q_{p+1}}{r^{6-p}}~~,
\label{gftb}
\\
g_{p+1,p+1} & = & \ldots~=~g_{99} 
\simeq 1 + \frac{p+1}{8}\,\frac{n^2}{\Delta_{m,n}}
~\frac{Q_{p+1}}{r^{6-p}}~~.
\nonumber
\end{eqnarray}
Using these expressions and assuming that the
complete solution of the (W,D$p$) bound
state can be written in terms of the harmonic
functions
$H$ and $H'$ defined in Eqs. (\ref{hr}) and (\ref{h1r})
(with $p$ replaced by $(p+1)$), we can
infer that the dilaton is 
\begin{equation}
{\rm e}^{\varphi} = {H'}^{{(3-p)}/{4}}~~,
\label{dilzz}
\end{equation}
and the metric is
\begin{eqnarray}
\label{metriczz}
ds^2 &=& -~H^{-1}\,{H'}^{{(p+1)}/{8}}
\left(dx^0\right)^2+H\,{H'}^{(p-7)/8}\left(dx^1
+A_0\,dx_0\right)^2
\nonumber \\
&+&
{H'}^{{(p-7)}/{8}}
\left[\left(dx^2\right)^2+\cdots+\left(dx^p\right)^2
\right] \nonumber \\
&+&
{H'}^{{(p+1)}/{8}}
\left[\left(dx^{p+1}\right)^2+\cdots+\left(dx^9\right)^2
\right] ~~,
\end{eqnarray}
where the Kaluza-Klein vector potential $A_0$ is
\begin{equation}
\label{a0zz}
 A_0 = \frac{m}{\Delta_{m,n}^{1/2}}~\left({H}^{-1}-1\right)~~.
\end{equation}
This solution agrees with the one presented in Ref.~\cite{LU2},
except that our Kaluza-Klein vector has a different sign and a
vanishing asymptotic value. 
For the R-R potentials, after performing the
T-duality, one proceeds as we discussed in
\secn{fdp} and obtains two
R-R (p+1)-forms, one of electric type and one of magnetic type,
whose expressions are similar to those of Eqs. (\ref{cpfin}) and
(\ref{cp1fin}) with obvious changes.

\end{document}